\documentclass[twocolumn]{aastex61}

\def\JB{{\rm Jy~beam^{-1}}}
\def\mJB{{\rm mJy~beam^{-1}}}

\def\mJ{{\rm mJy}}
\def\kms{{\rm km~s^{-1}}}
\def\Ms{M_{\sun}}

\def\II{{\rm I\hspace{-0.1em}I}}

\received{}
\revised{\today}
\accepted{}
\submitjournal{ApJ}

\shorttitle{Disk and envelope structures of L1527 IRS}
\shortauthors{Aso et al.}

\begin{document}


\title{ALMA OBSERVATIONS OF THE PROTOSTAR L1527 IRS: PROBING DETAILS OF THE DISK AND THE ENVELOPE STRUCTURES}

\correspondingauthor{Yusuke Aso}
\email{yusuke.aso@nao.ac.jp}

\author[0000-0002-8238-7709]{Yusuke Aso}
\affil{Department of Astronomy, Graduate School of Science, The University of Tokyo, 7-3-1 Hongo, Bunkyo-ku, Tokyo 113-0033, Japan}
\affil{Subaru Telescope, National Astronomical Observatory of Japan, 650 North A'ohoku Place, Hilo, HI 96720, USA}
\affil{Academia Sinica Institute of Astronomy and Astrophysics, P.O. Box 23-141, Taipei 10617, Taiwan}

\author{Nagayoshi Ohashi}
\affil{Subaru Telescope, National Astronomical Observatory of Japan, 650 North A'ohoku Place, Hilo, HI 96720, USA}
\affil{Academia Sinica Institute of Astronomy and Astrophysics, P.O. Box 23-141, Taipei 10617, Taiwan}

\author{Yuri Aikawa}
\affil{Center for Computational Sciences, University of Tsukuba, 1-1-1 Tennodai, Tsukuba, Ibaraki 305-8571, Japan}
\affil{Department of Astronomy, Graduate School of Science, The University of Tokyo, 7-3-1 Hongo, Bunkyo-ku, Tokyo 113-0033, Japan}

\author{Masahiro N. Machida}
\affil{Department of Earth and Planetary Sciences, Faculty of Sciences Kyushu University, Fukuoka 812-8581, Japan}

\author{Kazuya Saigo}
\affil{Chile Observatory, National Astronomical Observatory of Japan, Osawa 2-21-1, Mitaka, Tokyo 181-8588, Japan}

\author{Masao Saito}
\affil{Nobeyama Radio Observatory, Nobeyama, Minamimaki, Minamisaku, Nagano 384-1305, Japan}
\affil{SOKENDAI, Department of Astronomical Science, Graduate University for Advanced Studies}

\author{Shigehisa Takakuwa}
\affil{Department of Physics and Astronomy, Graduate School of Science and Engineering, Kagoshima University, 1-21-35 Korimoto, Kagoshima, Kagoshima 890-0065, Japan}
\affil{Academia Sinica Institute of Astronomy and Astrophysics, P.O. Box 23-141, Taipei 10617, Taiwan}

\author{Kengo Tomida}
\affil{Department of Earth and Space Science, Osaka University, Toyonaka, Osaka 560-0043, Japan}

\author{Kohji Tomisaka}
\affil{National Astronomical Observatory of Japan, Osawa, 2-21-1, Mitaka, Tokyo 181-8588, Japan}

\author{Hsi-Wei Yen}
\affil{European Southern Observatory, Karl-Schwarzschild-Str. 2, D-85748 Garching, Germany}



\begin{abstract}

We have newly observed the Class 0/I protostar L1527 IRS using the Atacama Large Millimeter/submillimeter Array (ALMA) during its Cycle 1 in 220 GHz dust continuum and C$^{18}$O ($J=2-1$) line emissions with a $\sim 2$ times higher angular resolution ($\sim 0\farcs5$) and $\sim 4$ times better sensitivity than our ALMA Cycle 0 observations. Continuum emission shows elongation perpendicular to the associated outflow, with a deconvolved size of $0\farcs 53\times 0\farcs 15$. C$^{18}$O emission shows similar elongation, indicating that both emissions trace the disk and the flattened envelope surrounding the protostar. The velocity gradient of the C$^{18}$O emission along the elongation due to rotation of the disk/envelope system is re-analyzed, identifying Keplerian rotation proportional to r$^{-0.5}$ more clearly than the Cycle 0 observations. The Keplerian-disk radius and the dynamical stellar mass are kinematically estimated to be $\sim 74$ AU and $\sim 0.45\ \Ms$, respectively. The continuum visibility is fitted by models without any annulus averaging, revealing that the disk is in hydrostatic equilibrium. The best-fit model also suggests a density jump by a factor of $\sim 5$ between the disk and the envelope, suggesting that disks around protostars can be geometrically distinguishable from the envelope from a viewpoint of density contrast. Importantly, the disk radius geometrically identified with the density jump is consistent with the radius kinematically estimated. Possible origin of the density jump due to the mass accretion from the envelope to the disk is discussed. C$^{18}$O observations can be reproduced by the same geometrical structures derived from the dust observations, with possible C$^{18}$O freeze-out and localized C$^{18}$O desorption.

\end{abstract}

\keywords{circumstellar matter --- stars: individual (L1527 IRS) --- stars: low-mass --- stars: protostars}



\section{INTRODUCTION} \label{sec:intro}
T Tauri stars are widely known to be associated with protoplanetary disks or T Tauri disks, which often show Keplerian rotation, as has been shown by interferometric observations in the last two decades \citep{gu.du1998,gu1999,si2000,qi2004,hu2009,ro2010}. Class 0/I protostars have also been, on the other hand, considered to be associated with protostellar disks, which are supposed to be precursors of T Tauri disks. Details of protostellar disks around protostars, such as their structures and dynamics, however, are poorly characterized because protostellar disks are deeply embedded in protostellar envelopes. The question of whether disks and envelopes in protostellar systems can be distinguished from each other by detailed observations of protostellar disks is therefore critical.

One promising way to achieve this goal is to distinguish them kinematically by inspecting their rotational motion, i.e., envelopes that often have infall motion with conserved angular momentum are expected to have rotation proportional to $r^{-1}$, while disks are expected to have Keplerian rotation proportional to $r^{-0.5}$. Initial attempts with such an approach have been successfully made recently, identifying a couple of disks with Keplerian rotation around protostars \citep{cho2010,ta2012,to2012,ye2013,mu2013,le2014}. Further attempts have been made more recently with higher angular resolutions ($\sim 100$ AU) and sensitivity, enabling us to identify even smaller, less bright disks around younger protostars \citep{ta2014,ye2014,oh2014,aso2015,le2016,ye2017}. Once envelopes and disks are kinematically distinguished, disk radii can also be kinematically estimated. Another important aspect to identify Keplerian rotation in disks around protostars is that it allows us to estimate dynamical masses of central protostars directly with no assumption. Such direct estimation of the dynamical mass enables us to examine whether or not infall motions around protostars are free fall, as has been assumed so far. L1527 IRS \citep{oh2014}, L1551 IRS 5 \citep{ch2014}, and TMC-1A \citep{aso2015} have infall motions that appear significantly slower than free fall velocities yielded by their dynamical masses, while L1489 IRS shows inflow at free fall velocity toward its Keplerian disk \citep{ye2014}.

These previous studies indicate that protostellar disks around protostars are kinematically similar to T Tauri disks in the sense that both disks show Keplerian rotation. It is not, however, clear whether protostellar disks are structurally similar to T Tauri disks as well. Physical structures of T Tauri disks, such as radial dependences of surface density and temperature have been well investigated, suggesting that their radial power-law profiles of surface density and temperature are well described as $\Sigma \propto r^{-p}$ with $p\sim 1$ regardless of exponential tails and $T\propto r^{-q}$ with $q\sim 0.5$, respectively. Their scale height described as $H\propto r^{h}$ was also investigated, and was found to be in hydrostatic equilibrium (HSEQ) with a power-law index of $h=1.25$. 
On the other hand, similar studies to investigate physical structures of protostellar disks have not been performed yet except for very limited cases 
with radial and vertical structures for the Butterfly Star \citep{wo2008}, L1527 IRS \citep{to2013}, TMC-1A \citep{aso2015}, four protostars in the NGC 1333 star forming region, and VLA 1623 \citep{pe2016} and without vertical structures for TMC-1A, TMC1, TMR1, L1536 \citep{ha2014}, and seven protostars in Perseus molecular cloud \citep{se2016}. Understanding physical structures of protostellar disks would be important to see whether disks and their surrounding envelopes can be geometrically distinguished from each other. If this is the case, we can have two independent ways (kinematically and geometrically) to identify disks around protostars. In addition, investigating physical structures of protostellar disks can help us to understand the disk formation process and also the process of evolution into T Tauri disks. In particular recent observations of T Tauri disks have revealed irregular structures such as spirals, gaps, and central holes, which might be related to planet formation. Tantalizing ring structures are discovered even in the disk around the class I/$\II$ young star HL Tau. It would be very important to understand when and how such structures are formed in these disks, and to answer this question, it is required to investigate structures of protostellar disks, which would be precursors of disks with irregular structures.

L1527 IRS (IRAS 04365+2557) is one of the youngest protostars, whose disk has been studied. L1527 IRS, located in one of the closest star-forming regions, the Taurus Molecular Cloud ($d=140$ pc), has bolometric luminosity $L_{\rm bol}=2.0\ L_{\odot}$ and bolometric temperature $T_{\rm bol}=44$ K \citep{kr2012}, indicating that L1527 IRS is a relatively young protostar. The systemic velocity of L1527 IRS in the local standard of rest (LSR) frame was estimated  to be $V_{\rm LSR}\sim 5.7\ \kms$ from C$^{18}$O $J=1-0$ observations with Nobeyama 45 m single-dish telescope \citep{oh1997}, while N$_{2}$H$^{+}$ $J=1-0$ observations with Five College Radio Astronomy Observatory (FCRAO) 14 m and Institut de Radioastronomie Millimetrique (IRAM) 30 m single-dish telescopes, estimated it to be $V_{\rm LSR}\sim5.9\ \kms$ \citep{ca2002,to2011}. We adopt $V_{\rm LSR}=5.8\ \kms$ for the systemic velocity of L1527 IRS in this paper, which is reasonable as will be shown later.

On a $\sim 30000$ AU scale, a bipolar outflow associated with this source was detected in the east-west direction by FCRAO single-dish observations in $^{12}$CO $J=1-0$ molecular line emission \citep{na2012} and by James Clerk Maxwell Telescope (JCMT) single-dish observations in $^{12}$CO $J=3-2$ molecular line emission \citep{ho1997}. Their results show that the blue and red robes of the outflows are on the eastern and western sides, respectively. On the other hand, inner parts ($\sim 8000$ AU scale) of the outflow mapped with Nobeyama Millimeter Array (NMA) in $^{12}$CO $J=1-0$ show the opposite distribution, i.e., stronger blueshifted emission on the western side and stronger redshifted emission on the eastern side  \citep{ta1996}.  
Mid-infrared observations toward L1527 IRS with Spitzer Space Telescope shows bright bipolar scattered light nebulae along the outflow axis in the $\sim 20000$ AU scale \citep{to2008}. They fitted a protostellar envelope model to near- and mid-infrared scattered light images and spectral energy distribution (SED). As a result the inclination angle of the envelope around L1527 IRS was estimated to be $i=85^{\circ}$, where $i=90^{\circ}$ means the edge-on configuration. In addition \citet{oy2015} found that the western side is closer to observers.
This inclination angle is consistent with a disk-like structure of dust highly elongated along the north-south direction, which was spatially resolved for the first time in 7-mm continuum emission by the Very Large Array (VLA) \citep{lo2002}.
By expanding the studies by \citet{to2008}, \citet{to2013} fitted a model composed of an envelope and a disk to (sub)millimeter continuum emissions and visibilities observed with SMA and CARMA as well as infrared images and SED. Their best-fitting model suggests a highly flared disk structure ($H\propto R^{1.3}$, $H=48$ AU at $R=100$ AU) with a radius of 125 AU. This study has geometrically distinguished the protostellar disk and envelope around L1527 IRS, although they were not kinematically distinguished from one another. 

The first interferometric observations of the envelope surrounding L1527 IRS in molecular line emission were reported by \citet{oh1997}, identifying an edge-on flattened envelope elongated perpendicularly to the associated outflow, in their C$^{18}$O $J=1-0$ map obtained with NMA at an angular resolution of $\sim 6\arcsec$. It was found that kinematics of the envelope can be explained with dynamical infall motion $(\sim 0.3\ \kms)$ and slower rotation ($\sim 0.05\ \kms$) at 2000 AU. Its mass infalling rate was also estimated to be $\dot{M}\sim 1\times 10^{-6}\ \Ms \, {\rm yr}^{-1}$. 
Higher-resolution ($\sim 1\arcsec$) observations using Combined Array for Research in Millimeter Astronomy (CARMA) in $^{13}$CO $J=2-1$ line were carried out toward this source by \citet{to2012}. They measured emission offsets from the central protostar at each channel and fitted the position-velocity data with a kinematical model using the LIne Modeling Engine (LIME) by assuming Keplerian rotation. According to their best-fit result, the mass of L1527 IRS was estimated to be $M_{*}=0.19\pm 0.04\ \Ms$ and the disk radius was also estimated to be 150 AU. It should be noted, however, that no other kinds of rotation, such as the one conserving its angular momentum, were compared with the observations in the work. They attempted later to compare their rotation curve with rotation laws conserving angular momentum, which did not change their original conclusion \citep{to2013pro}. 

In order to investigate the rotational velocity around L1527 IRS without assuming Keplerian rotation, a radial profile of the rotational velocity $V_{\rm rot}$ was measured by \citet{ye2013} from their SMA observations in C$^{18}$O $J=2-1$ line with the angular resolution of $4\farcs 2 \times 2\farcs 5$. In their analysis, the rotation profiles are derived from Position-Velocity (PV) diagrams cutting along a line perpendicularly to the outflow axis. The rotation profile of L1527 IRS was measured at $r\gtrsim 140$ AU to be $V_{\rm rot}\propto r^{-1.0\pm 0.2}$, which is clearly different from Keplerian rotation $V_{\rm rot}\propto r^{-1/2}$.
Further investigation of the rotation profile around L1527 IRS was performed with much higher sensitivity as well as higher angular resolution provided by Atacama Large Millimeter/submillimeter Array (ALMA) \citep{oh2014}. The rotation profile obtained in C$^{18}$O $J=2-1$ at a resolution of $\sim 0\farcs9$ mostly shows velocity inversely proportional to the radius being consistent with the results obtained by \citet{ye2013}, while it also suggests a possibility that the profile at $\lesssim 54$ AU can be interpreted as Keplerian rotation with a central stellar mass of $\sim 0.3\ \Ms$. In addition, infall velocity in the envelope is found to be slower than the free fall velocity yielded by the expected central stellar mass \citep{oh2014}.

In this paper we report new ALMA Cycle 1 observations of L1527 IRS in C$^{18}$O $(J=2-1)$ and 220 GHz continuum, with a $\sim 2$ times higher angular resolution and a $\sim 4$ times higher sensitivity as compared with our previous ALMA cycle 0 observations, which allow us to give a much better constraint on the rotation profile of the disks and the envelope, and also their geometrical structures. Our observations and data reduction are described in Section \ref{sec:obs}. In Section \ref{sec:res}, we present the continuum and molecular-line results. In Section \ref{sec:ana}, we analyze rotation velocity measured by the C$^{18}$O line and perform $\chi ^{2}$ fitting to explain the continuum visibility using a model. In Section \ref{sec:disc}, we investigate the validity and consistency of the model that reproduces the observations the best. We present a summary of the results and our interpretation in Section \ref{sec:conc}.

\section{ALMA OBSERVATIONS AND DATA REDUCTION} \label{sec:obs}

We observed our target, L1527 IRS, during Cycle 1 using ALMA on 2014 July 20. The observations were composed of two tracks in the same day with a separation of $\sim 40$ minutes. Each track was $\sim 30$ minutes including overhead. J0510$+$1800 was observed as the passband, gain, and flux calibrator for the former track and J0423-013 was observed as the flux calibrator for the latter track. Thirty-four antennas were used in the first track, while one antenna was flagged in the latter track. The antenna configuration covers projected baseline length from 17 to 648 m (13-474 k$\lambda$ in $uv$-distance at the frequency of C$^{18}$O $J=2-1$). This minimum baseline resolves out more than 50\% of the flux when a structure is extended more than $7\farcs 1$ \citep{wi.we1994}, corresponding to $\sim990$ AU at a distance of L1527 IRS. The coordinates of the map center during the observations were $\alpha {\rm (J2000)}=04^{\rm h}39^{\rm m}53\fs90,\ \delta {\rm (J2000)}=26^{\circ}03\arcmin 10\farcs00$. C$^{18}$O $J=2-1$ line and 220 GHz continuum emission in Band 6 were observed for 6.9$+$6.7$=$14 minutes (on source). To achieve high velocity resolution for molecular line observations, we configured the correlator in Frequency Division Mode for two spectral windows. Each spectral window has 3840 channels covering 234 MHz bandwidth. Emission-free channels in the lower side band are used to make the continuum map centered at 220 GHz. The total bandwidth of the continuum map is $\sim234$ MHz. 

We performed self-calibration for the continuum observations using tasks ($clean$, $gaincal$, and $applycal$) in Common Astronomy Software Applications (CASA), and the obtained calibration table for the continuum observations was applied to the C$^{18}$O observations. The self-calibration has improved the rms noise level of the continuum map by a factor of 2-3, while the noise level of the C$^{18}$O map has been improved by less than a few percent. The noise level of the C$^{18}$O map was measured in emission-free channels. 

All the mapping process was carried out with CASA. 
Because the original map center during the observations was not coincident with the continuum peak position estimated from 2D Gaussian fitting in the $uv$-domain by $0\farcs54$, the phase center of the observed visibilities was shifted from the original phase center with $fixvis$ in CASA, making the map center of the resultant maps in this paper the same as the continuum peak position. 
The visibilities were Fourier transformed and CLEANed. In this process we adopted superuniform weighting with $npixel=2$ and binned two frequency channels; the resultant frequency resolution in this paper is 122 kHz, corresponding to 0.17 $\kms$ in the velocity resolution at the frequency of C$^{18}$O $J=2-1$. We set a $12\arcsec \times 12\arcsec$ area centered on the map center as a CLEAN box with a threshold of $3\sigma$. The synthesized beam sizes of the CLEANed maps are $0\farcs 50\times 0\farcs 40 $ for the C$^{18}$O line, and $0\farcs47\times 0\farcs37$ for the continuum emission.  
The parameters of our observations mentioned above and others are summarized in Table \ref{tab:obs}.

\begin{deluxetable}{c|cc}
\tablecaption{Summary of the ALMA observational parameters \label{tab:obs}}
\tablehead{
\colhead{Date} & \multicolumn{2}{c}{2014.Jul.20}\\
\colhead{Target} & \multicolumn{2}{c}{L1527 IRS}\\
\colhead{Coordinate center} & \multicolumn{2}{c}{R.A. (J2000)=4$^{\rm h}$39$^{\rm m}$53$^{\rm s}\!\!$.9}\\
\colhead{} & \multicolumn{2}{c}{Dec. (J2000)=26$^{\circ }03\arcmin 10\farcs 0$}\\
\colhead{Projected baseline length} & \multicolumn{2}{c}{17.4 - 647.6 m}\\
\colhead{Primary beam} & \multicolumn{2}{c}{28\farcs 6}\\
\colhead{Passband calibrator} & \multicolumn{2}{c}{J0510$+$1800}\\
\colhead{Flux calibrator} & \multicolumn{2}{c}{J0423$-$013, J0510$+$1800}\\
\colhead{Gain calibrator} & \multicolumn{2}{c}{J0510$+$1800}}
\colnumbers
\startdata
 & Continuum & C$^{18}$O $J=2-1$\\
\hline
Frequency & 219.564200 & 219.560358\\
Synthesized beam (P.A.) & $0\farcs 47 \times 0\farcs 37\ (-0.4^{\circ})$ & $0\farcs 50\times 0\farcs 40\ (3.1^{\circ})$\\
Velocity resolution & 234 MHz & 0.17 $\kms$\\
1$\sigma$ & 0.2 $\mJB$ & 2.6 $\mJB$\\
\enddata
\end{deluxetable}

\section{RESULTS} \label{sec:res}
\subsection{220 GHz Continuum} \label{sec:cont}

\begin{figure}[ht!]
\figurenum{1}
\epsscale{1}
\plotone{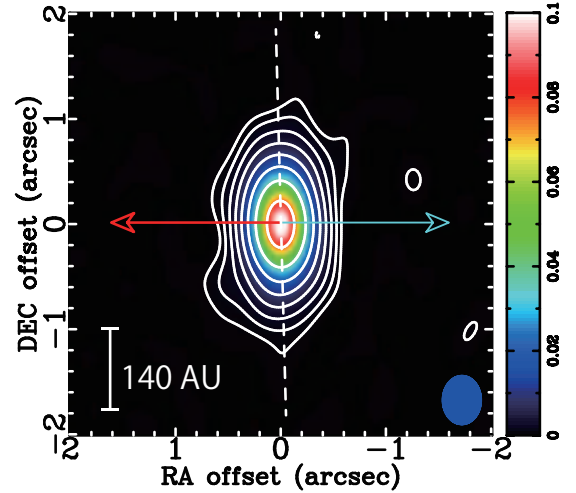}
\caption{Continuum emission map of L1527 IRS. Contour levels are $-3,3,6,12,24,\dots \times \sigma$, where 1$\sigma$ corresponds to $0.2\ \mJB$. A blue-filled ellipse at the bottom right corner denotes the ALMA synthesized beam; $0\farcs 47\times 0\farcs 37,\ {\rm P.A.}=-0.4^{\circ}$. The elongation direction ($1.5^{\circ}$) is shown with a white dashed line. Blue and red arrows show the direction of the molecular outflow (east-west) from single-dish observations toward L1527 IRS in $^{12}$CO $J=1-0$ \citep{na2012}.
\label{fig:ci}}
\end{figure}

Figure \ref{fig:ci} shows the 220 GHz continuum emission in L1527 IRS observed with ALMA. Strong compact continuum emission is detected. The emission is clearly elongated in the north-south direction and shows weak extensions to the northwest and the southeast. The $6\sigma$ contour in Figure \ref{fig:ci} shows a full width of $\sim 2\arcsec=280$ AU along the north-south direction. Its deconvolved size derived from a 2D Gaussian fitting is $531\pm 2\ {\rm mas}\times 150\pm2\ {\rm mas},\ {\rm P.A.}=1.5^{\circ}\pm 0.2^{\circ}$. This major-axis direction is almost perpendicular to the direction of the associated outflow, indicating that the continuum emission traces a dust disk and/or a flattened dust envelope around L1527 IRS. Compared with the synthesized beam size ($0\farcs 47\times 0\farcs 37,\ {\rm P.A.}=-0.4^{\circ}$), the major axis of the emission is longer than the beam size and thus spatially resolved, which is consistent with the previous observations at a higher angular resolution \citep[$\sim 0\farcs 35$;][]{se2015}\citep[$\sim 0\farcs3$;][]{to2013}. The aspect ratio, 0.28, is a half of (i.e. thinner than) the ratio reported by \citet{oh2014} with lower angular resolutions than this work. The peak position is also measured from the Gaussian fitting to be $\alpha (2000)=04^{\rm h}39^{\rm m}53\fs 88,\ \delta (2000)=+26^{\circ}03\arcmin 09\farcs 55$, which is consistent with previous measurements \citep{ye2013,oh2014}. We define this peak position and the major-axis direction as the central protostellar position of L1527 IRS and the orientation angle of its dust disk/envelope respectively in this paper. The peak intensity and the total flux density of the emission derived from the Gaussian fitting are $101.4\pm 0.2\ \mJB$ and $164.6\pm 0.5\ \mJ$, respectively, while the total flux density is $176\ \mJ$ when measured in the whole region of Figure \ref{fig:ci}. By assuming that the dust continuum emission is optically thin and dust temperature is isothermal, total mass can be calculated with the total flux density \citep{an2005}. The total fluxes derived above correspond to a mass of $M_{\rm gas}\sim 0.013\ \Ms$ by assuming a dust opacity of $\kappa(220\ {\rm GHz})=0.031\ {\rm cm}^{2}\,{\rm g}^{-1}$ \citep{to2013}, a dust temperature of 30 K \citep{to2013}, and a standard gas to dust mass ratio, g/d, of 100. 

\subsection{C$^{18}$O J=2-1} \label{sec:18}

\begin{figure}[ht!]
\figurenum{2}
\epsscale{1}
\plotone{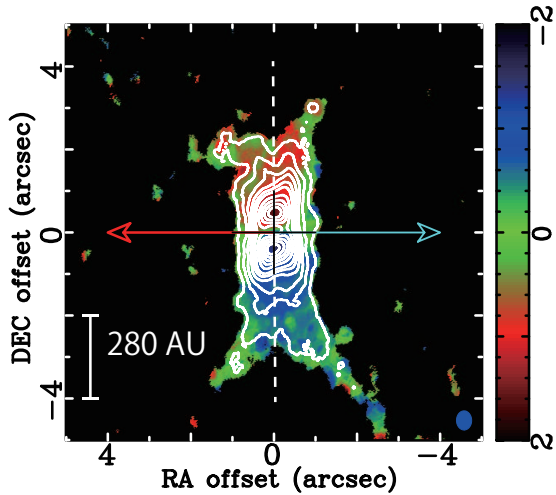}
\caption{Integrated intensity map (moment 0; white contours) and mean velocity map (moment 1; color) of the C$^{18}$O $J=2-1$ emission in L1527 IRS, where the velocity is relative LSR velocity with respect to the systemic velocity $V_{\rm LSR}=5.8\ \kms$. Contour levels of the integrated intensity map are from $5\sigma$ to $30\sigma$ in steps of $5\sigma$ and then in steps of $10\sigma$, where $1\sigma$ corresponds to $2.3\ \mJB \, \kms$. A central black plus sign shows the position of the central protostar (continuum emission peak). A blue-filled ellipse at the bottom right corner denotes the ALMA synthesized beam; $0\farcs 50\times 0\farcs 40,\ {\rm P.A.}=3.1^{\circ}$. Blue/red arrows and a white dashed line show the direction of the molecular outflow and the major-axis direction of the continuum emission, respectively, as shown in Figure \ref{fig:ci}.
\label{fig:18m}}
\end{figure}

The C$^{18}$O $J=2-1$ emission was detected above $3\sigma$ level at the relative velocity range from $-3.3$ to $3.2\ \kms$
in the LSR frame with respect to the systemic velocity $V_{\rm LSR}=5.8\ \kms$. Figure \ref{fig:18m} shows the total integrated intensity (moment 0) map in white contours and the intensity-weighted-mean velocity (moment 1) map in color; both are derived from the above velocity range with $3\sigma$ cutoff.
The moment 0 map overall shows an elongated structure perpendicular to the outflow axis, centered at the protostellar position. In more detail, lower contours ($\sim 3$-$6\sigma$) show extensions to north-northeast, north-northwest, south-southeast, and south-southwest. The moment 0 map also shows two local peaks on the northern and southern sides of the central protostar with a separation of $\sim 1\arcsec$. This double peak is due to a ``continuum subtraction artifact''; the continuum emission was subtracted even at channels where the C$^{18}$O emission has low contrast with respect to the continuum emission and is resolved out by the interferometer. Subtraction of the continuum thus results in negative intensity at the protostellar position. Regardless of the double peak, the map was fitted with single 2D Gaussian to measure the overall structure of the C$^{18}$O emission; a deconvolved size of the C$^{18}$O emission is estimated to be $2\farcs 17\pm 0\farcs 04 \times 0\farcs 88\pm 0\farcs 02$, with ${\rm P.A.}=-1.8^{\circ}\pm 0.7^{\circ}$. Peak integrated intensity and total flux measured in the whole region of Figure \ref{fig:18m} are $0.20\ \JB \, \kms$ and $2.2\ {\rm Jy} \, \kms$. The moment 1 map shows a velocity gradient in the north-south direction, which is perpendicular to the outflow axis.
The morphology of C$^{18}$O emission indicates that it traces a flattened gas envelope and/or a gas disk around L1527 IRS and thus the velocity gradient seen in the C$^{18}$O emission is mainly due to their rotation, as already suggested by \citet{oh2014} and \citet{ye2013}.
Because the C$^{18}$O emission shows a more complicated structure than the continuum emission, we assume the orientation angle of the gas disk/envelope to be the same as that of the dust disk/envelope in this paper.

\begin{figure*}[ht!]
\figurenum{3}
\plotone{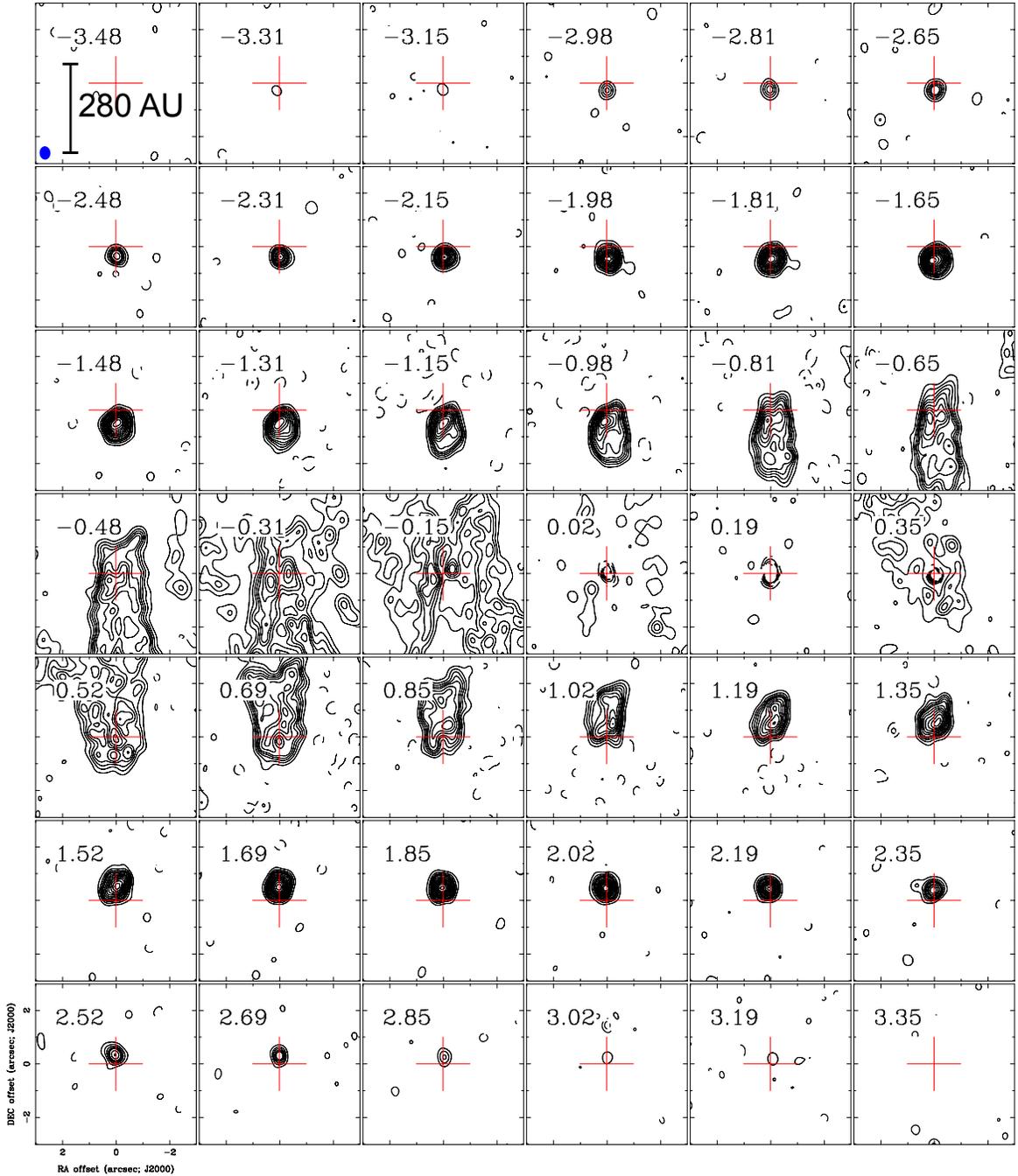}
\caption{Channel maps of the C$^{18}$O $J=2-1$ emission in L1527 IRS. Contour levels are from $3\sigma$ to $15\sigma$ in steps of $3\sigma$ and then in $5\sigma$, where $1\sigma$ corresponds to $2.6\ \mJB$. A central red plus sign in each panel shows the position of the central protostar (continuum emission peak). A blue-filled ellipse in the top left panel denotes the ALMA synthesized beam; $0\farcs 50\times 0\farcs 40,\ {\rm P.A.}=3.1^{\circ}$. Relative LSR velocity with respect to the systemic velocity $V_{\rm LSR}=5.8\ \kms$ is shown at the top left corner of each panel.
\label{fig:18c}}
\end{figure*}

Figure \ref{fig:18c} shows channel maps of the C$^{18}$O emission, which enable us to investigate velocity structures in more detail. In higher blue- and redshifted velocities ($|V|\gtrsim 1.6\ \kms$), emissions show overall circular shapes and their sizes at $3\sigma$ level are smaller than $\sim 1\farcs 5$. The emission peaks are located on the southern side in the blueshifted range while on the northern side in the redshifted range, making a velocity gradient from the south to the north as seen in Figure \ref{fig:18m}. In a middle velocity range ($0.4\lesssim |V|\lesssim 1.5\ \kms$), more complicated structures can be seen. For example, at $-1.15,\ -0.98,\ 0.85$, 1.02 $\kms$, the emissions appear to composed of a strong compact ($\sim 1\arcsec$) structure close to the protostar and a more extended ($>2\arcsec$) structure, resulting in a plateau structure. The extended emissions are located mainly on the southern side in the blueshifted range while mainly on the northern side in the redshifted range. Furthermore some blueshifted channels ($-0.65$ and $-0.48\ \kms$) show an extension from the protostar to the northwest. In the other lower velocities, emissions are strongly resolved out and negative emission can be seen from 0.02 to $0.35\ \kms$. This redshifted negative emission is due to a continuum subtraction artifact with an extended infalling envelope around L1527 IRS as \citet{oh2014} confirmed using a infalling envelope$+$Keplerian disk model with radiative transfer calculation. The higher angular resolution than \citet{oh2014} can make the negative emission deeper and result in a double peak in Figure \ref{fig:18m} that did not appear in \citet{oh2014}. 
Other recent observations reported that spatial thickness of the envelope decreases outward from a radius of $\sim 150$ AU by a factor of two in CCH ($N=4-3$, $J=9/2-5/2$, $F=5-4$ and $4-3$) line emission \citep{sa2017}. No such structure, however, can be confirmed in our results in C$^{18}$O emission, which is considered to be a better tracer of the overall column density and H$_{2}$ gas distribution. The CCH emission has a higher critical density than the C$^{18}$O emission by three orders of magnitude and thus traces only dense regions in the envelope.

\begin{figure*}[ht!]
\figurenum{4}
\gridline{
\fig{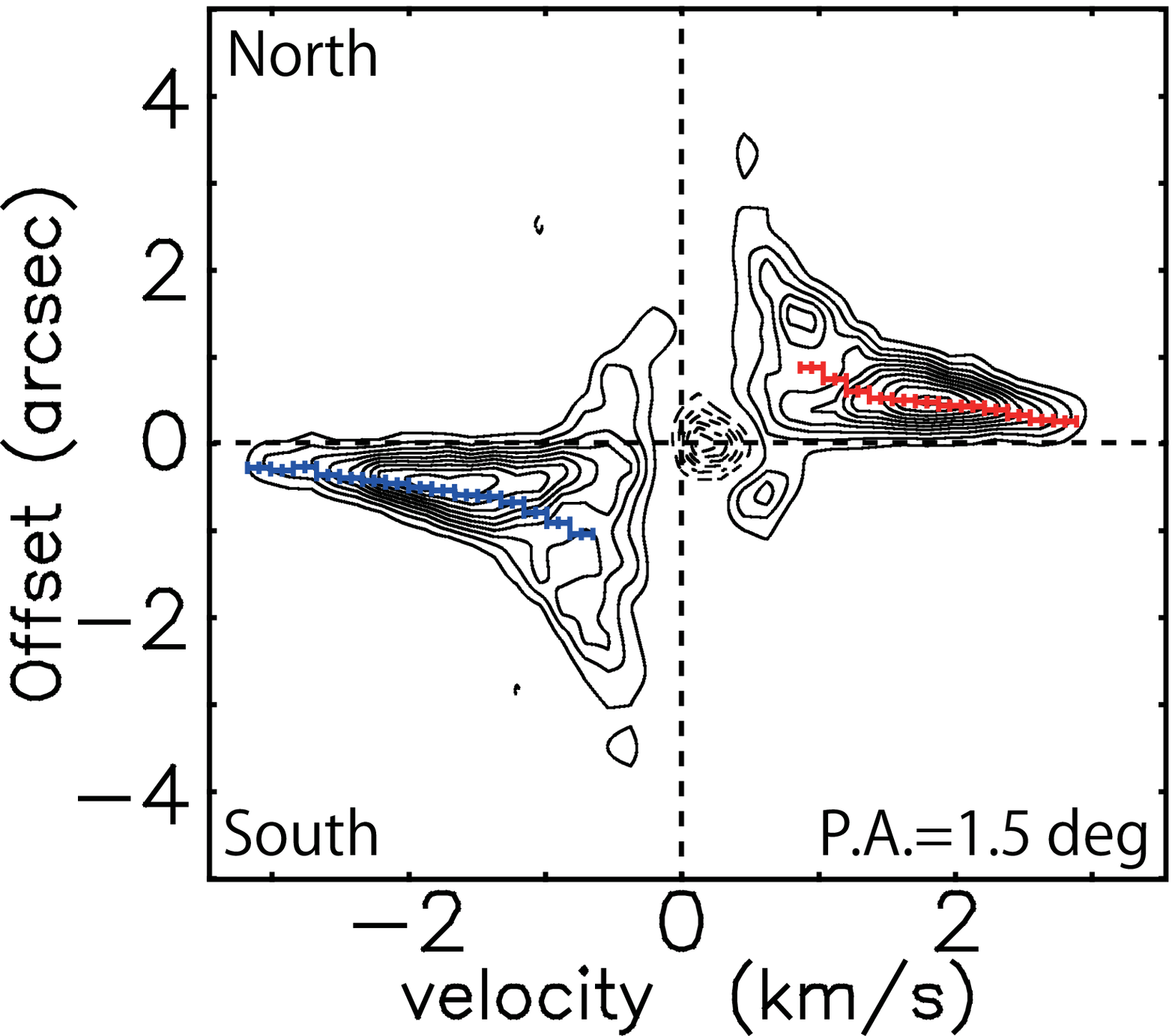}{0.5\textwidth}{(a)}
\fig{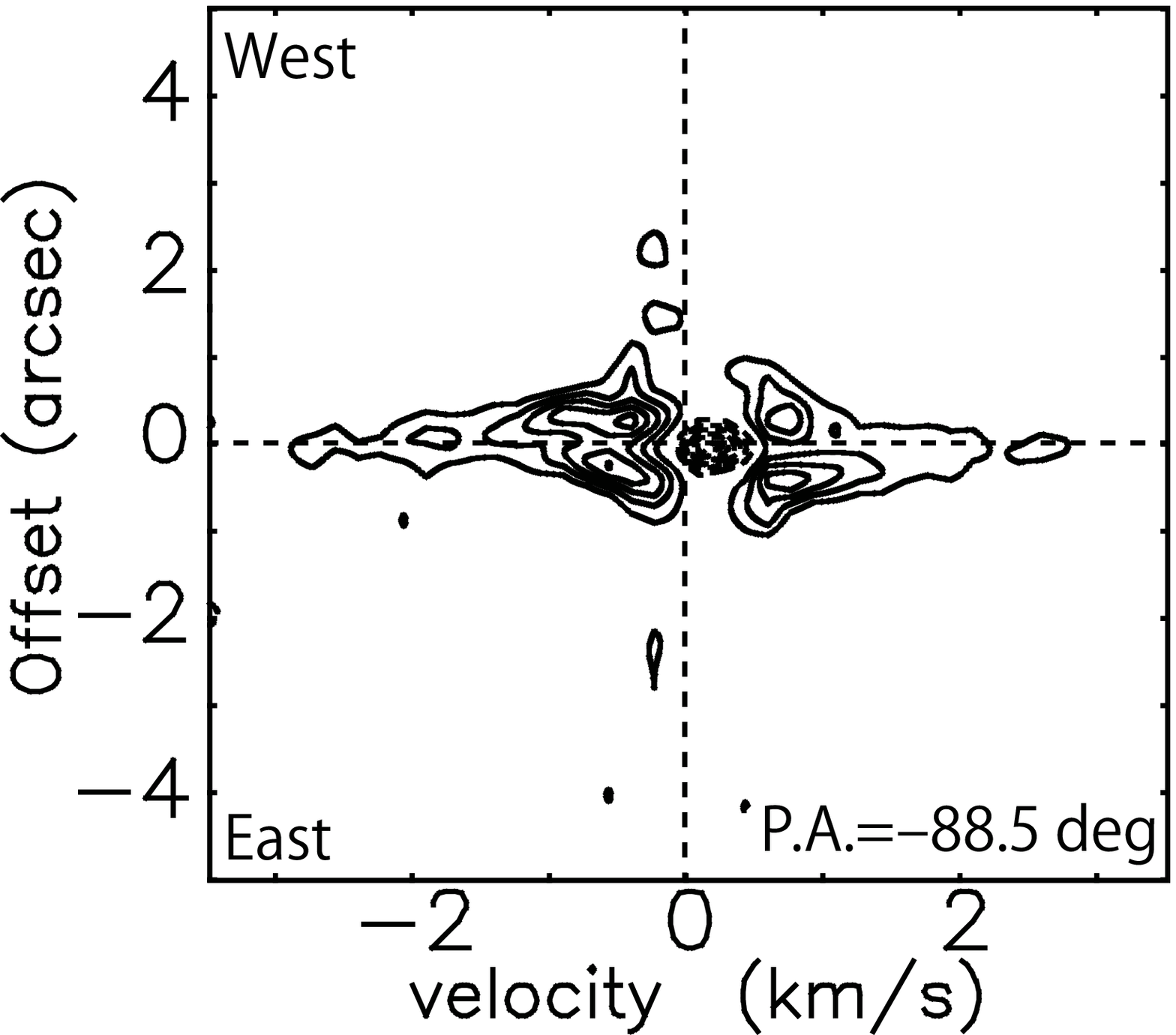}{0.5\textwidth}{(b)}
}
\caption{Position Velocity diagrams of the C$^{18}$O $J=2-1$ emission in L1527 IRS along (a) the major axis and (b) the minor axis of the continuum emission (the major axis corresponds to the white dashed line in Figure \ref{fig:ci}, ${\rm P.A.}=1.5^{\circ}$). The width of cut is one pixel, $0\farcs02$. These PV diagrams have the same angular and velocity resolutions as those of the channel maps shown in Figure \ref{fig:18c}. Contour levels are $5\sigma$ spacing from $3\sigma$, where $1\sigma$ corresponds to $2.6\ \mJB$. Central vertical dashed lines show the systemic velocity and central horizontal dashed lines show the protostellar position. Blue and red points with error bars in panel (a) are mean positions derived along the position (vertical) direction at each velocity.
\label{fig:18pv}}
\end{figure*}

In the moment 1 map and channel maps, L1527 IRS shows a velocity gradient from the south to the north. The PV diagrams along the major and minor axes are considered to represent a velocity gradient due to rotation and radial motion, respectively, toward disk-like structures as discussed in previous work. Figure \ref{fig:18pv}a and b show PV diagrams of the C$^{18}$O emission cutting along the major and minor axes, respectively. The overall velocity gradient from the south to the north can be confirmed in the PV diagram along the major axis. In more detail, the so-called ``spin up'' rotation can also be seen in $|V|\gtrsim 1.5\ \kms$, that is, an emission peak at a velocity channel is closer to the central position at higher velocity. We will analyze the dependence of rotational velocity on radial distance from the central position in Section \ref{sec:rp}. In $0.6\lesssim |V|\lesssim 1.4\ \kms$, a strong compact emission and a more extended emission appear to be superposed, which corresponds to the plateau structures seen in the channel maps (Figure \ref{fig:18c}).
In the PV diagram along the minor axis, there are four strong peaks in the western redshifted and blueshifted components, and the eastern blueshifted and redshifted ones in $|V|\lesssim 1.5\ \kms$ while the emission is mainly concentrated on the central position in the higher velocity range. The western blueshifted component extends to higher velocities than the eastern blueshifted one, and also the eastern redshifted component extends to higher velocities than the western redshifted one. These extensions can be interpreted as a small velocity gradient from the west to the east, which was detected in observations in CS $J=5-4$ line emission and considered to be due to infalling motion in the protostellar envelope by \citet{oy2015}. 

\section{ANALYSIS} \label{sec:ana}
\subsection{Rotation Profile} \label{sec:rp}
\begin{figure}[ht!]
\figurenum{5}
\epsscale{1}
\plotone{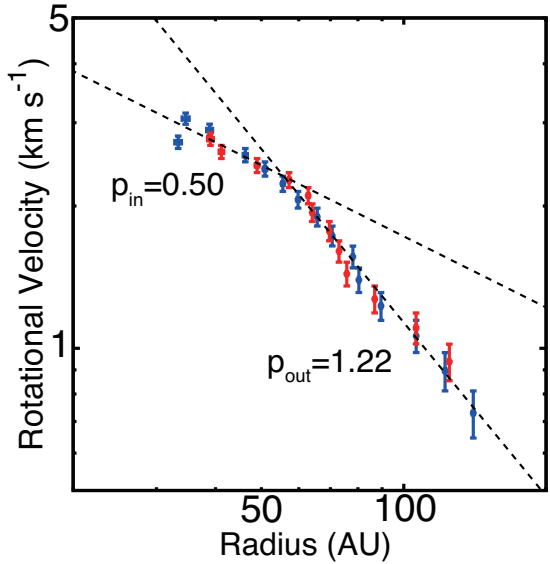}
\caption{Mean positions of the PV diagram along the major axis plotted on a $\log R-\log V$ plane. The ordinate is not deprojected. Blue and red points show blueshifted and redshifted mean positions in Figure \ref{fig:18pv}a, respectively. Dashed lines show the best-fit lines with a double power law.
\label{fig:18log}}
\end{figure}

In the previous section we identified rotation in the C$^{18}$O gaseous component tracing either a flattened envelope, disk, or both. In this section the radial profile of the rotation will be investigated with the PV diagram along the major axis (Figure \ref{fig:18pv}) so as to characterize the nature of the observed rotation. 
The method used in this section to obtain the radial profile of rotational velocity is based on the analyses presented by \citet{ye2013} and also explained by \citet{aso2015} in detail. The representative position at each velocity channel of the PV diagram along the major axis is measured as the intensity-weighted 1D mean position, $x_{m}(v)=\int xI(x,v)dx /\int I(x,v)dx$. Pixels having intensity more than $5\sigma$ are used to calculate the sum. The error bar of each representative position is also derived by considering propagation of errors. The derived representative positions are overlaid with error bars on the PV diagram along the major axis (Figure \ref{fig:18pv}) and also plotted in a $\log R-\log V$ plane (Figure \ref{fig:18log}). The abscissa of Figure \ref{fig:18log} is calculated from the offset position in the PV diagram by assuming that the distance of L1527 IRS is 140 pc. Figure \ref{fig:18log} shows a clear negative correlation that the rotational velocity is higher at the position closer to the central protostar, i.e., differential rotation. Furthermore the $\log R-\log V$ diagram in Figure \ref{fig:18log} exhibits two different linear regimes with a break radius of $\sim 60$ AU.
The data points in the $\log R-\log V$ plane are, therefore, fitted by a double power-law function with four free parameters: inner and outer power-law indices $p_{\rm in}$ and $p_{\rm out}$, respectively, and a break point $(R_{b},V_{b})$ \citep[see Equation (1) in][]{aso2015}. The best-fit parameter set is $(R_{b},V_{b},p_{\rm in},p_{\rm out})=(56\pm 2\ {\rm AU}, 2.31\pm 0.07\ \kms ,0.50\pm 0.05, 1.22\pm 0.04)$, giving a reasonable reduced $\chi ^{2}$ of 1.6. For comparison, $\chi ^{2}$ fitting with a single power function is also performed, where $V_{b}$ is fixed at $2.31\ \kms$. The best-fit parameter set for this case is $(R_{b},p)=(49.5\pm 0.3\ {\rm AU},0.88\pm 0.01)$, giving reduced $\chi ^{2}=4.0$. These reduced $\chi^{2}$ suggest that the radial profile of rotational velocity is characterized by the double power function better than any single power function.
In addition, to examine whether the LSR velocity of L1527 IRS we adopt, $5.8\ \kms$, is reasonable, we also carried out fitting including the systemic velocity as another free parameter, which was fixed to be $5.8\ \kms$ above, confirming that the adopted systemic velocity is the most reasonable to fit the $\log R-\log V$ diagram.

The best-fit inner power-law index is almost equal to Keplerian rotation law ($p=1/2$), suggesting that the inner/higher-velocity component of the C$^{18}$O emission traces a Keplerian disk. In fact, the Keplerian rotation was marginally detected in \citet{oh2014}. 
On the other hand, the best-fit outer power-law index is roughly equal to that of rotation conserving its angular momentum ($p=1$), which is steeper than the Keplerian rotation law, and thus suggests that the rotation in the envelope around the disk cannot support material against the gravity yielded by the central protostar. This is consistent with the fact that the envelope is in infalling motion, as was reported by \citet{oh2014}. 
The best-fit result is quite consistent with that obtained in our previous Cycle 0 observations of L1527 IRS \citep{oh2014}, while the higher angular resolution and sensitivity of the Cycle 1 observations has enabled us to sample twice as many data points within the break radius in the $\log R-\log V$ diagram as the previous work, making the break in the rotation profile more definite with more precise measurements of the radius of the break and the inner power-law index.
The radius of the identified Keplerian disk can be estimated from the best-fit break radius. Note that low angular resolutions along the minor axis of disks cause emission within the radius to be measured at a given velocity channel and the emission moves the representative position inward at the channel, as pointed out by \citet{aso2015}. This underestimation strongly affects in the case of edge-on configurations, such as L1527 IRS, because such configurations make the angular resolution even lower along the minor axis. We thus take into account the underestimation. In their notation, the correction factor is a constant, 0.760, when Keplerian disk radius $R_{o}$ is smaller than $\lesssim 480$ AU with the inclination angle $i=85^{\circ}$ and the angular resolution $\theta \sim 0\farcs 5=70$ AU.
With this correction factor taken into account, the Keplerian disk radius is estimated to be $\sim 74$ AU. From the best-fit break velocity together with this disk radius, the central protostellar mass $M_{*}$ of L1527 IRS and a specific angular momentum $j$ at the outermost radius of the Keplerian disk can also be estimated to be $M_{*}\sim 0.45\ \Ms$ and $j\sim 8.3\times 10^{-4}\ \kms \, {\rm pc}$, respectively, where the inclination angle is assumed to be $i=85^{\circ}$ \citep{to2013,oy2015}.

\subsection{Structures of the Keplerian Disk} \label{sec:cv}

As shown in the previous section, a Keplerian disk has been kinematically identified by the C$^{18}$O results. This disk around the protostar L1527 IRS seems to be kinematically quite similar to those around T Tauri stars \citep{gu.du1998,si2000,ro2010}. A tantalizing question is whether this disk is also geometrically similar to those around T Tauri stars. Because this disk is almost edge-on, it is also possible to investigate the vertical structures of the disk. In this subsection geometrical structures of the Keplerian disk are investigated.

\subsubsection{Continuum Visibility distribution} \label{sec:cvd}
\begin{figure*}[ht!]
\figurenum{6}
\plottwo{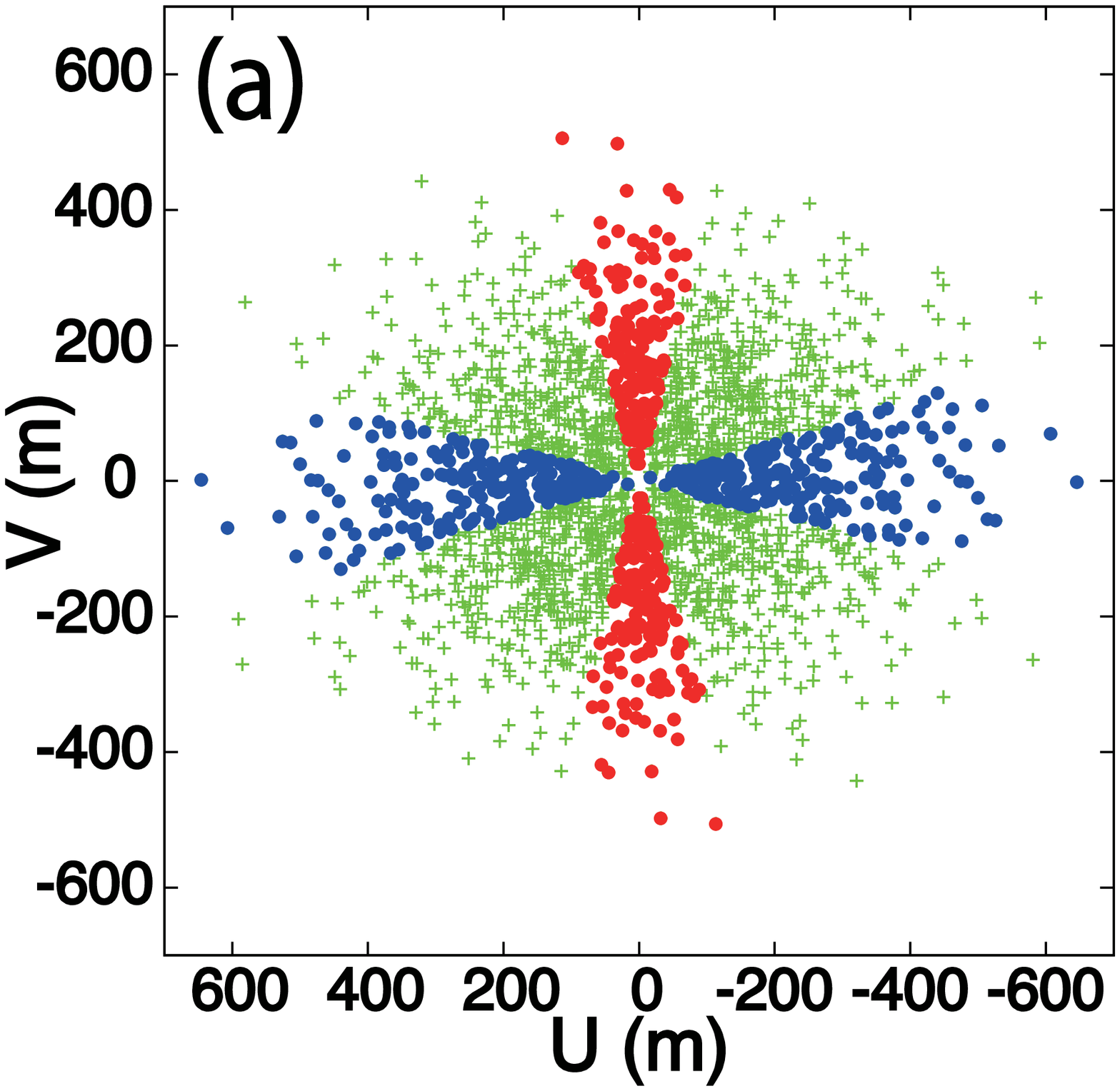}{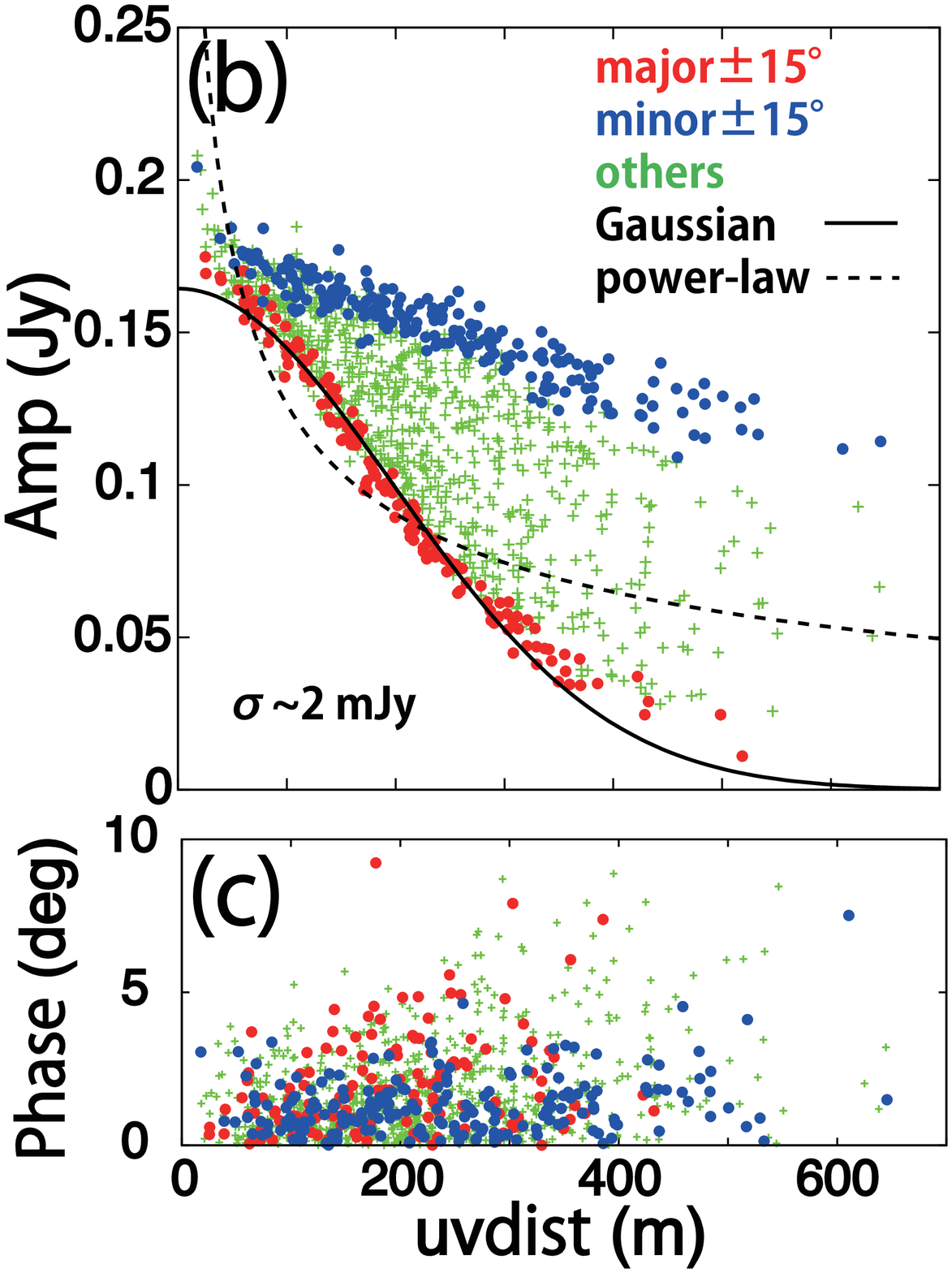}
\caption{Continuum visibility averaged over scans. The two observational tracks are not averaged together. 1 m corresponds to $0.73\ {\rm k}\lambda$ at the observed frequency. (a) Data points on the $uv$-plane. Both of each conjugate pair are plotted.  (b) Distribution of the visibility amplitude. An error bar of the amplitude for each point is $\sim 2\ {\rm mJy}$. (c) Distribution of the visibility phase. Only one of each conjugate pair with a positive phase is plotted. Red and blue circles denote data points near the major- and minor-axis directions, respectively; in other words, ${\rm Arctan}(U/V)=1.5^{\circ}\pm 15^{\circ},\ -88.5^{\circ}\pm 15^{\circ}$, respectively. Green crosses denote the other data points.
Solid and dashed curves are the best-fit Gaussian and power-law profiles, respectively, to the data points near the major axis (red points). 
\label{fig:cuvd}}
\end{figure*}

To investigate the disk structures, we have performed direct model fitting to the observed continuum visibility data, which is free from any non-linear effects associated with interferometric imaging. Figure \ref{fig:cuvd} shows distributions of the continuum visibilities in three panels, where red and blue points denote the same groups; the red points are located near the major axis ($\pm 15^{\circ}$ on the $uv$-plane) while the blue points are located near the minor axis. 
Note that all the visibilities of each baseline in each track for $\sim30$ minutes are averaged in these plots. That is why any trajectory due to Earth's rotation is not drawn on the $uv$-plane (Figure \ref{fig:cuvd}a). It should be stressed that although all the visibilities of each baseline are averaged, no further azimuthal average has been done in our analysis as is obvious in Figure \ref{fig:cuvd}a. This is because information on structures that are not spherically symmetric such as disks is missed with azimuthally averaged visibilities unless the disks are face-on, as explained below in more detail.

Figure \ref{fig:cuvd}b exhibits a trend that the visibility amplitudes are higher at shorter $uv$-distances. The total flux density $176\ \mJ$ measured in the image space (Figure \ref{fig:ci}) appears consistent with an amplitude at zero $uv$-distance, which can be derived from visual extrapolation of the amplitude distribution.
Figure \ref{fig:cuvd}b also exhibits the data points appearing to scatter more widely at longer $uv$-distances. This is clearly not due to the error of each data point, which is $\sim 2\ {\rm mJy}$ but due to the structures of the continuum emission. In more detail, blue and red points have the highest and lowest amplitudes at each $uv$-distance, respectively. Because visibility is derived by Fourier transforming an image, these distributions of the blue and red points indicate that structures of the continuum emission are largest along the major axis and smallest along the minor axis in the image domain, which is consistent with the image of the continuum emission shown in Figure \ref{fig:ci}. Similarly the scattering of the green points between the blue and red points is due to structures along directions at different azimuthal angles.
In addition, the data points near the major axis (red points) are also compared with two simple functions in Figure \ref{fig:cuvd}b: Gaussian and power-law profiles. Best-fit profiles are $0.16\ {\rm Jy}\ \exp (-4\ln 2 (\beta /234\ {\rm m})^{2})$ and $0.13\ {\rm Jy}\ (\beta /100\ {\rm m})^{-0.48}$, respectively, where $\beta$ denotes the $uv$-distance. The power-law profile cannot explain the observations at all. Although the Gaussian profile matches the observations better, this profile is not necessarily realistic for circumstellar disk structures \citep[][and references therein]{ha2014} and also the comparison shows systematic deviation; the points are higher than the Gaussian profiles in $<100$ m and $>300$ m while they are lower in $\sim 200$ m.
Figure \ref{fig:cuvd}c exhibits that most phases of the visibility are smaller than $\lesssim 5^{\circ}$, which corresponds to $\lesssim 0\farcs 04$ where $uv$-distance is larger than 100 m. This indicates that emission is centered at the protostellar position at most spatial frequencies. Red points and blue points appear to be well mixed in Figure \ref{fig:cuvd}c, which means that the distribution of phases in the azimuthal direction is roughly uniform. 

As Figure \ref{fig:cuvd}b clearly demonstrates, the analysis of visibility without azimuthal average in a $uv$-plane is quite powerful for investigating spatially resolved not-spherically symmetric structures such as disks except for face-on cases. 
No such analysis with sufficient signal-to-noise ratio has been done in previous studies. Note that a few studies attempted to use 2D distributions in model fitting \citep{pe2016}. Those studies, however, showed only azimuthally averaged visibilities deprojected by considering inclination angles, but lacked the signal-to-noise ratio to perform comparable exploration to the 2D visibilities, making it impossible to evaluate how good their fittings were in 2D $uv$-space. Exceedingly high sensitivity as well as high angular resolution of ALMA allows us to perform such data analyses.

\subsubsection{Model fitting}
\label{sec:fit}
Our analysis of the disk structures is performed by $\chi^{2}$ fitting of models to the continuum visibilities shown in Figure \ref{fig:cuvd}b. It should be noted that the full size of the continuum emission at $6\sigma$ level in Figure \ref{fig:ci}, $\sim 280$ AU (see Section \ref{sec:cont}) \footnote{The $6\sigma$ size is referred here to see the extension of the whole continuum emission, which is much larger than the FWHM derived from Gaussian fitting.} is twice the disk size expected from the radius kinematically estimated in Section \ref{sec:rp}. This suggests that the continuum emission arises not only from the disk but also from the envelope. Hence our models should include envelope structures as well as disk structures. Because the envelope around L1527 IRS shows a flattened morphology, the model we use in this section is based on a standard disk model \citep[e.g.,][]{du1994} but modified to express a flattened dust envelope as well as a dust disk, as described below. We used the code described in \citet{oh2014}. The model includes 12 parameters summarized in Table \ref{tab:mod}. The radial dependence of temperature $T(R)$ and scale height $H(R)$ are described as $T(R)=T_{1}(R/1\ {\rm AU})^{-q}$ and $H(R)=H_{1}(R/1\ {\rm AU})^{h}$, respectively. This means that the scale height in our model is not assumed to be in HSEQ. To express dust disk and dust envelope structures, the radial dependence of surface density $\Sigma (R)$ is described by a combination of inner and outer profiles formulated as
\begin{eqnarray}
\Sigma (R)=\frac{(2-p)M_{\rm disk}}{2\pi \left( R_{\rm out}^{2-p} -R_{\rm in}^{2-p}\right) }R^{-p}\times
\left\{
\begin{array}{c}
1\ \ (R\leq R_{\rm out})\\
S_{\rm damp}\ \ (R>R_{\rm out})
\end{array}
\right.
,
\label{eq:sd}
\end{eqnarray}
where $M_{\rm disk}$ is a disk mass (mass within $R_{\rm out}$) determined from g/d$=100$ and $S_{\rm damp}$ is a damping factor of the surface density for the outer dust envelope. The model has no envelope when $S_{\rm damp}$ is zero, while the model has no density jump when $S_{\rm damp}$ is unity. $R_{\rm out}$ is the outer radius of the disk defined as the boundary between the disk and the envelope. In our model, mass density distribution $\rho(R,z)$ is determined from the scale height $H(R)$ and the surface density $\Sigma (R)$ as $\Sigma/(\sqrt{2\pi}H)\exp (-z^{2}/2H^{2})$ within the outermost radius of 1000 AU. Our model adopts the same power-law index of density for both disk and envelope, which is consistent with theoretical simulations by \citet{ma2010}. Some of the parameters, including the power-law index of the temperature distribution, are fixed as shown in Table \ref{tab:mod}. Radio observations are considered to be not sensitive to temperature distribution because most observed continuum emissions are optically thin, which makes it hard to constrain temperature distribution by mm-continuum observations. Therefore, we fixed the temperature distribution in our model, referring a total luminosity derived in infrared wavelengths \citep{to2008,to2013}, which are more sensitive to temperature than radio observations. In addition, because our angular resolution is not high enough to resolve vertical structures of the disk, the radial temperature profile we adopted is vertically isothermal. We confirmed that our profile provides representative temperature of the 2D temperature distribution derived by \citet{to2013} using a self-consistent radiative transfer model \citep{wh2003} at each radius on 10-100 AU scales. The inclination angle of the dust disk/envelope is fixed at $i=85^{\circ}$ and the eastern side is on the near side for the observers \citep{oy2015}. The other six quantities are free parameters ($M_{\rm disk},R_{\rm out},p,S_{\rm damp},H_{1},h$). When radiative transfer were solved in 3D space to produce a model image, the following condition and quantities were assumed: local thermodynamic equilibrium (LTE), g/d$=100$, and a dust opacity of $0.031\ {\rm cm}^{2}\, {\rm g}^{-1}$ calculated from a opacity coefficient of $\kappa (850\ \mu {\rm m})=0.035\ {\rm cm}^{2}\, {\rm g}^{-1}$ and an opacity index of $\beta =0.25$ \citep{to2013}.

After a model image was calculated from the radiative transfer, model visibility was obtained by synthetic observations through the CASA tasks $simobserve$ and $listvis$. In the synthetic observations, the phase center was set at the center of the model image and the orientation (P.A.) was assumed to be the same as that of the observed continuum emission. Following the two observational tracks, we performed the synthetic observations without artificial noise with the same antenna configurations as the observations. Then model visibilities were derived at the same points on the $uv$-plane as the observations. Using the model visibility and the observed continuum visibility, reduced $\chi ^{2}$ was calculated to evaluate the validity of each model. Without azimuthally averaging visibilities, all 1089 data points were used to calculate the reduced $\chi ^{2}$ defined as
\begin{eqnarray}
\chi^{2}_{\nu}=&&\frac{1}{\sigma ^{2}(2N_{\rm data}-N_{\rm par}-1)}\sum _{i}\left[ \left( {\rm Re} V_{i}^{\rm obs}-{\rm Re} V_{i}^{\rm mod}\right)^{2}\right. \nonumber \\
&&+\left. \left( {\rm Im} V_{i}^{\rm obs}-{\rm Im} V_{i}^{\rm mod}\right)^{2}\right],
\end{eqnarray}
where $\sigma$, $V_{i}^{\rm obs}$, $V_{i}^{\rm mod}$, $N_{\rm data}$, and $N_{\rm par}$ are the standard deviation of noise in the observed visibility amplitude, the observed visibility, model visibility, the number of data points, and the number of free parameters, respectively; $N_{\rm data}=1089$ and $N_{\rm par}=6$ as mentioned above. $N_{\rm data}$ is multiplied by two because each visibility includes two independent values, real and imaginary parts. Using the distribution of reduced $\chi ^{2}$ in the parameter space, the uncertainty of each parameter is defined as the range of the parameter where the reduced $\chi ^{2}$ is below the minimum plus one $(=6.6)$ when all parameters are varied simultaneously. We also used the Markov Chain Monte Carlo method to find the minimum $\chi ^{2}$ efficiently.

\begin{figure*}[ht!]
\figurenum{7}
\gridline{
\fig{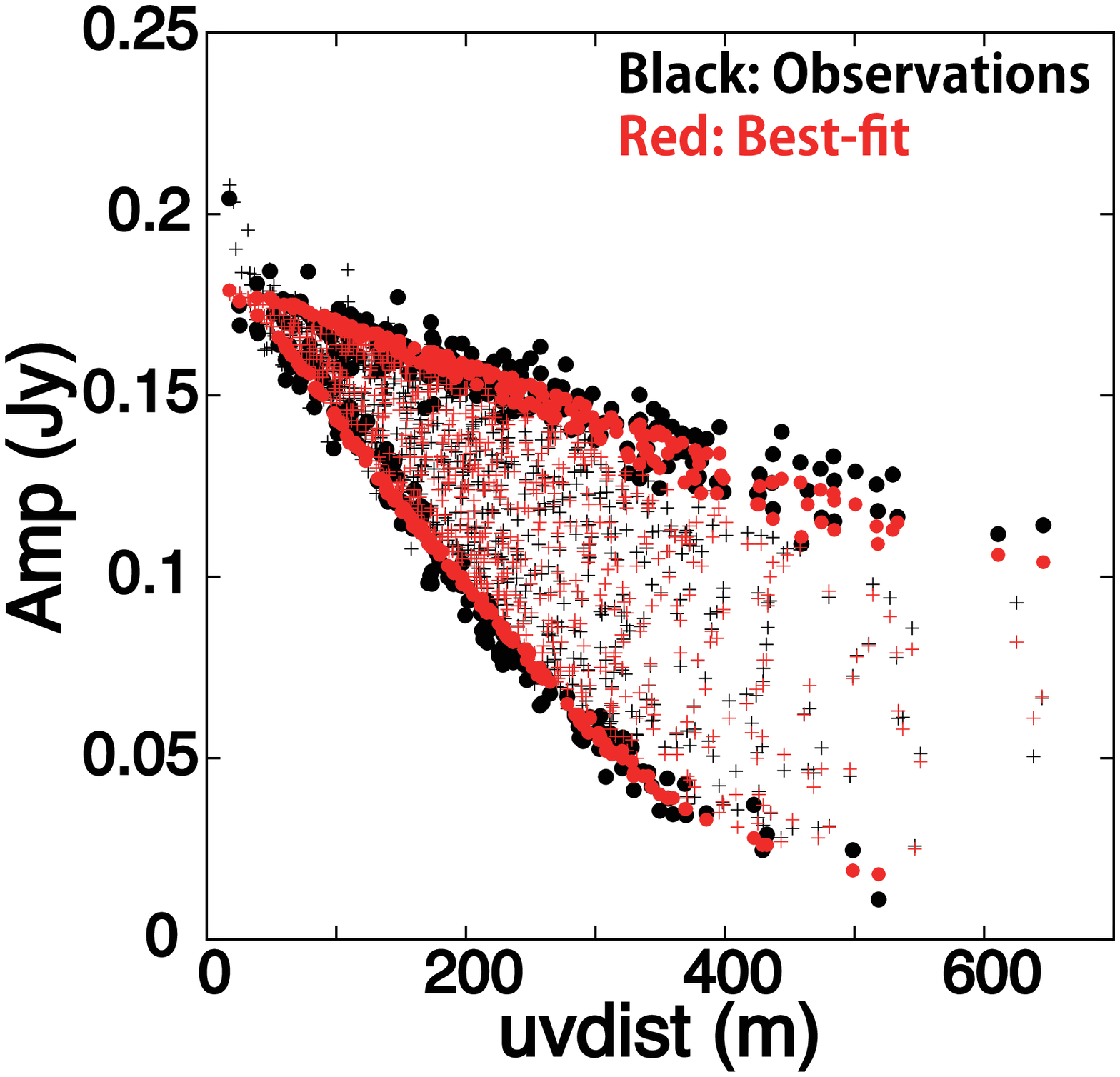}{0.33\textwidth}{(a)}
\fig{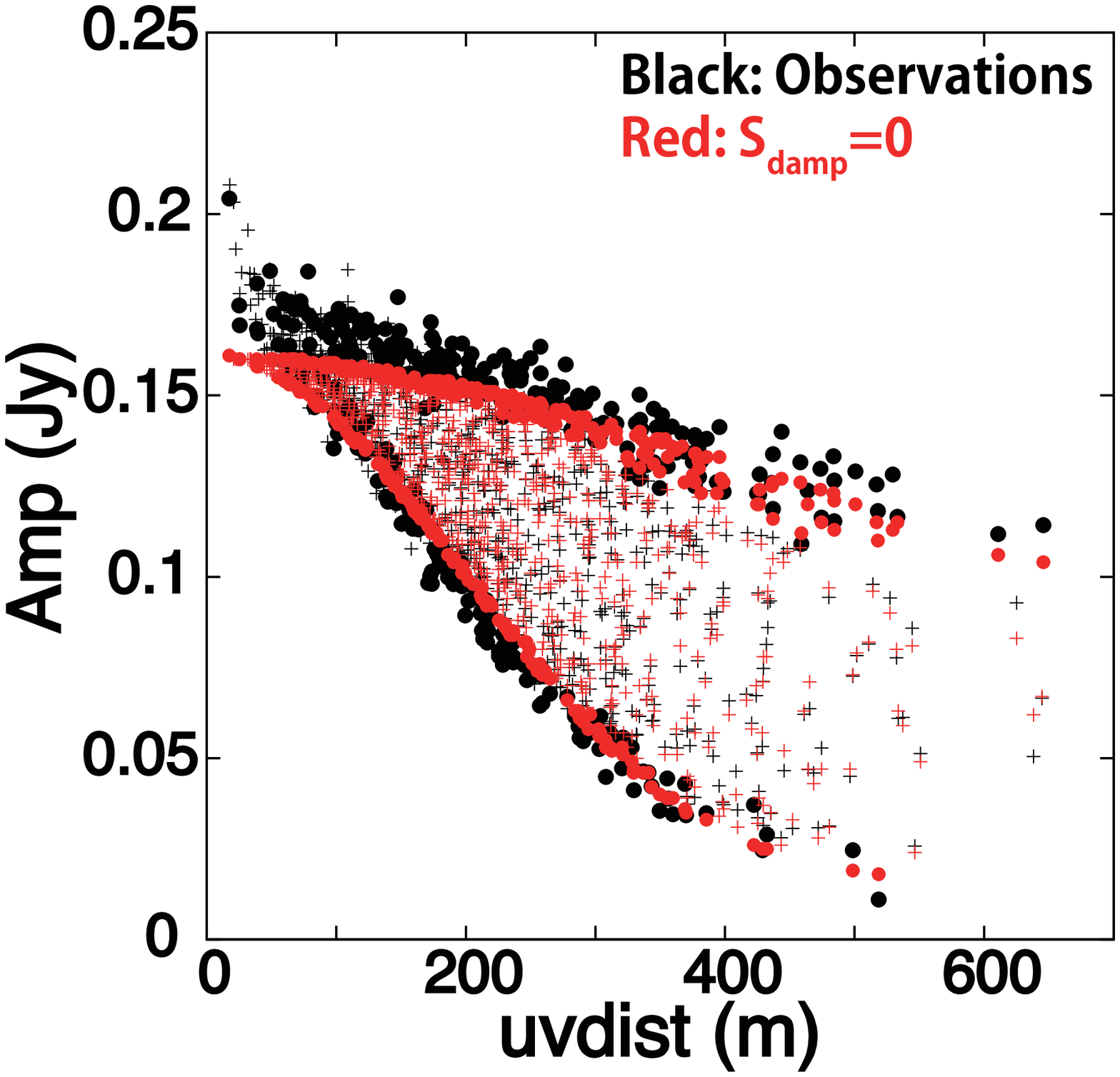}{0.33\textwidth}{(b)}
\fig{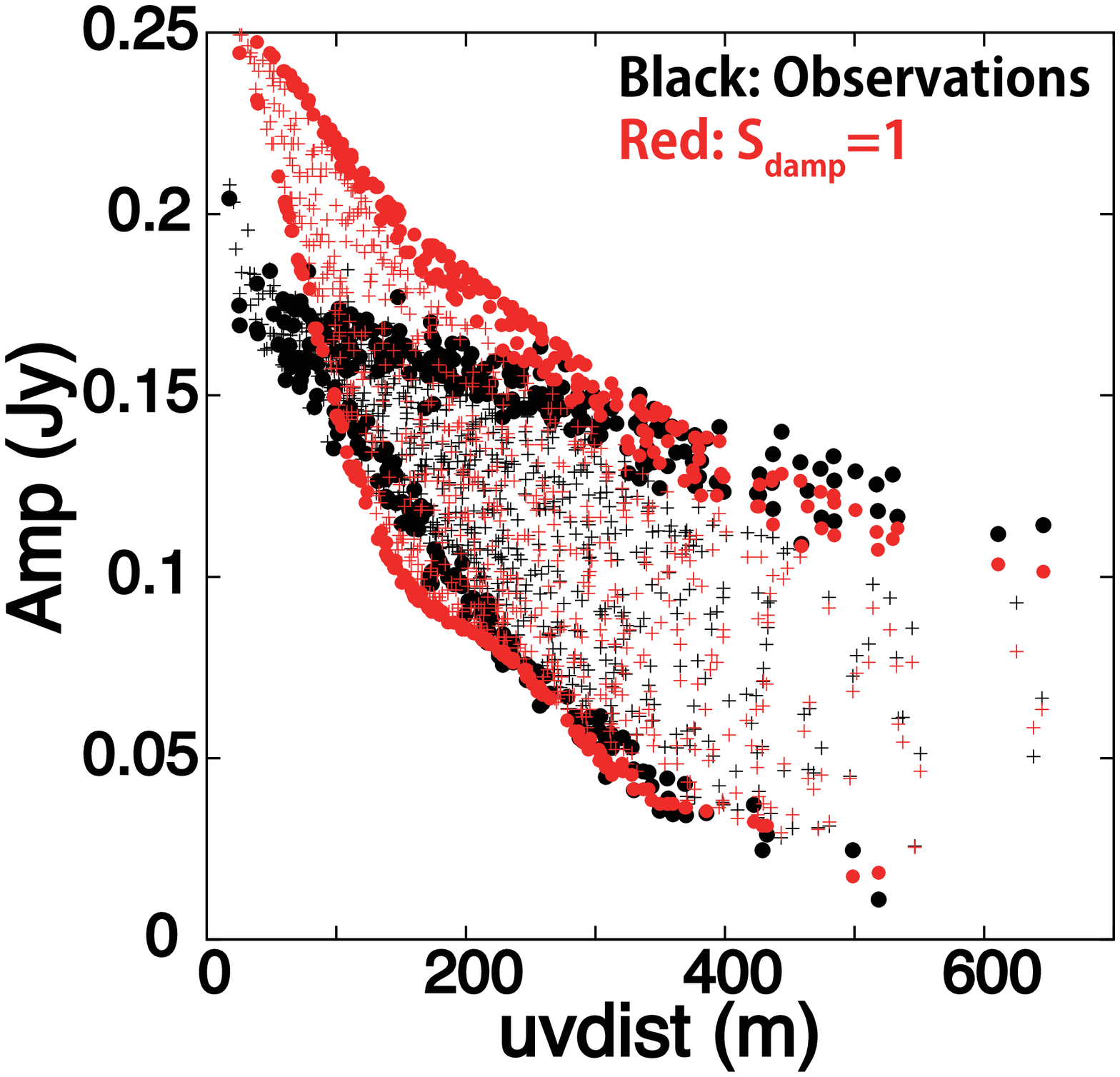}{0.33\textwidth}{(c)}
}
\caption{The observed continuum visibility (black circles and crosses) and (a) the best-fit model, (b) the model with $S_{\rm damp}=0$, and (c) the model with $S_{\rm damp}=1$, denoted with red circles and crosses. For the models in panel (b) and (c), parameters except for $S_{\rm damp}$ are fixed at those of the best-fit model. The observations are the same plot as Figure \ref{fig:cuvd}(b) except for the color, and thus the circles denote the data points near the major- and minor-axis directions while crosses denote the other data points for both the observations and models.
\label{fig:ccv}}
\end{figure*}

\begin{figure}[ht!]
\figurenum{8}
\gridline{
\fig{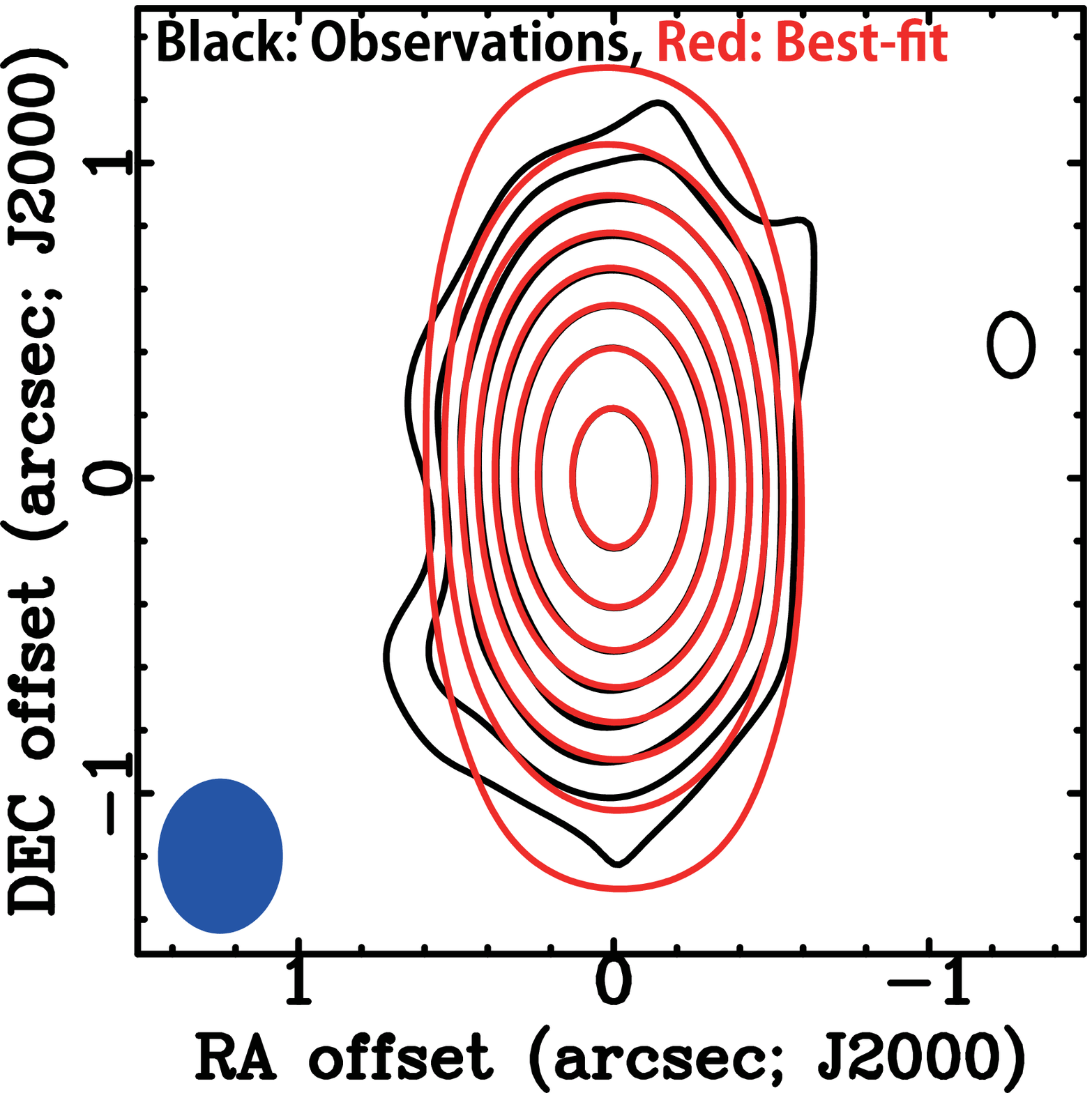}{0.25\textwidth}{(a)}
\fig{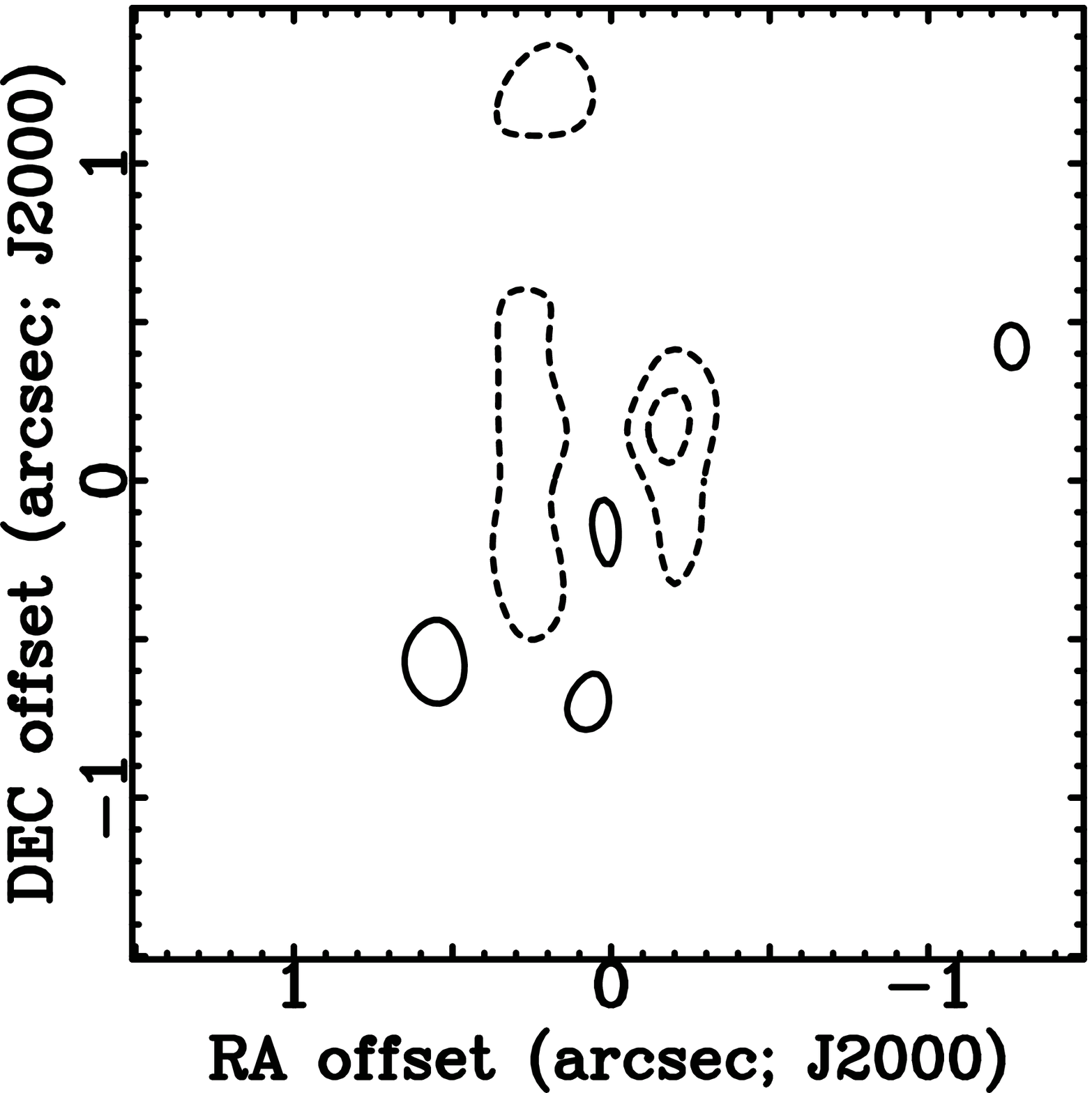}{0.25\textwidth}{(b)}
}
\caption{(a) The observed continuum image (black contours) and the best-fit model (red contours). (b) The residual obtained by subtracting the best-fit model from the observations. Contour levels are $-3,3,6,12,24,\dots \times \sigma$ for panel (a) and $-3,3,6,9,\dots \times \sigma$ for panel (b), where $1\sigma$ corresponds to $0.2\ \mJB$. A blue-filled ellipse at the bottom right corner in panel (a) denotes the ALMA synthesized beam; $0\farcs 47\times 0\farcs 37,\ {\rm P.A.}=-0.4^{\circ}$. The spatial scale is different from Figure \ref{fig:ci}.
\label{fig:cci}}
\end{figure}

Figure \ref{fig:ccv}a represents a comparison of the observed continuum visibility with our best-fit model showing that the best-fit model overall reproduces the observations. The reduced $\chi ^{2}$ of the best-fit model is 5.6, which corresponds to a residual of $\sim 2.4\sigma$ on average in the $uv$-space. It is also confirmed that the best-fit model can reproduce the observations in an image space, as shown in Figure \ref{fig:cci}. Note that the model image in Figure \ref{fig:cci}a was not made through the synthetic observation using CASA but was simply made based on the best-fit parameters in Table \ref{tab:mod} with convolution using the synthesized beam of our observations. This is because the synthetic observations using CASA produce a beam that is slightly different from the actual observations. Synthetic observation is not very crucial for this comparison, but convolution with the beam exactly same as the synthesized beam used in the actual observations is more crucial because the model image is relatively compact as compared to the synthesized beam. The residual shown in Figure \ref{fig:cci}b was obtained by subtracting the best-fit model, in the image space, from the observations, indicating that almost no significant residual can be seen in the image space. This comparison demonstrates that when the residual in the $uv$-space is small, the image-space residual derived from the analysis above is also small.

The parameters of the best-fit model are summarized in Table \ref{tab:mod}. The damping factor $S_{\rm damp}=0.19^{+0.03}_{-0.09}$, which is not zero but significantly smaller than unity, suggests that the observed continuum emission arises from both a dust disk and a dust envelope with a significant jump of the column density between them. A much larger or smaller value than 0.19 cannot explain the observed visibilities as shown in Figure \ref{fig:ccv}b and \ref{fig:ccv}c. The former and the latter show models with $S_{\rm damp}=0$ and 1, respectively, where the other parameters are the same as those of the best-fit model. In Figure \ref{fig:ccv}b and \ref{fig:ccv}c visibility amplitude of models at longer $uv$-distance ($\gtrsim 300$ m) appears to be similar to the observations in both cases, whereas visibility amplitude of the models with $S_{\rm damp}=0$ and 1 are lower and higher than the observations, respectively, at shorter $uv$-distance ($\lesssim 300$ m). One might wonder whether a jump of dust opacity, as well as surface density, could also explain the observations. It is observationally confirmed, however, that the opacity index $\beta$ does not change on scales of 100 to 1000 AU, based on observed dependence of spectral index on $uv$-distance \citep{to2013}, and possible uncertainty of $\Delta \beta$, $\sim 0.2$, adds only $\sim 8$\% to the relative uncertainty of $S_{\rm damp}$. Further discussion on the significant jump of the surface density between the dust disk and envelope will be presented in Section \ref{sec:disc}. The best-fit value of $R_{\rm out}=84^{+16}_{-24}$ AU is close to the Keplerian disk radius kinematically derived from the C$^{18}$O results; the discrepancy between the two values are within their errors. This result suggests that the dust disk identified geometrically as density contrast by this model fitting is consistent with the kinematically identified gaseous Keplerian disk. The power-law index of the surface density, $p=1.7^{+0.1}_{-0.3}$, is a bit steeper than the typical value for T Tauri disks, $\sim 1.0$ \citep{an.wi2007,hu2008,an2010} while a similarly steep $p$ has been found toward a few other protostars as well \citep[e.g.,][]{ye2014,aso2015}. The power-law index of the scale height, $h=1.2^{+0.1}_{-0.1}$, corresponds to that for HSEQ ($h=1.25$) within the error range, where the temperature distribution is assumed to be vertically isothermal. To examine whether the best-fit model is indeed in HSEQ, the scale height of the disk at a certain radius can be compared directly with that in HSEQ. The scale height of the best-fit model at $R=R_{\rm out}$ is calculated to be $\sim 20$ AU while the scale height of a disk in HSEQ is estimated to be $\sim 15$ AU at the same radius when the central stellar mass is $M_{*}=0.45\ \Ms$ as kinematically estimated from the rotation profile (see Section \ref{sec:rp}), the temperature is 44 K as estimated from the temperature profile we used in the model, and a mean molecular weight is $2.37\ m_{\rm H}$. This comparison suggests that the disk around L1527 IRS is most probably in HSEQ. Note that the scale height in HSEQ also depends on vertical temperature distribution; it can be higher if temperature is higher in upper layers of disks. \citet{to2013} suggested a highly flared disk around L1527 IRS, which has a scale height of 38 AU at $R_{\rm out}$. We consider that their estimated scale height is higher than ours because their observations included infrared wavelength, which may be traced by scattered light making the disk geometrically thick in appearance, or because smaller grains traced by infrared wavelength are in upper layers than larger grains traced by mm wavelength. In addition, the scale height and the index derived by our best-fit model are relatively large and steep, respectively, when compared with protoplanetary disks in Ophiuchus star forming region \citep{an2010}.

The best-fit model provides us a comparison between the disk around L1527 IRS and T Tauri disks from a geometrical point of view. The disk around L1527 IRS has a mass and a radius similar to those of T Tauri disks. On the other hand, the power-law index of the disk surface density, the scale height, and the power-law index of the scale height are possibly steeper, larger, and steeper, respectively, than those of T Tauri disks.

\begin{deluxetable*}{ccccccc}
\tablecaption{Fixed and Free Parameters of the Model Fitting \label{tab:mod}}
\tablehead{\colhead{Fixed} & \colhead{$i$} & \colhead{$R_{in}$} & \colhead{$T_{1}$} & \colhead{$q$} & \colhead{$\kappa (220\ {\rm GHz})$} & \colhead{g/d}\\
 & \colhead{$85^{\circ}$} & \colhead{0.1 AU} & \colhead{403.5 K} & \colhead{0.5} & \colhead{$0.031\ {\rm cm}^{2}\, {\rm g}^{-1}$} & \colhead{100}}
\startdata
Free & $M_{\rm disk}$ & $R_{\rm out}$ & $p$ & $S_{\rm damp}$ & $H_{1}$ & $h$\\
Best & $1.3^{+0.3}_{-0.4}\times 10^{-2}\ \Ms$ & $84^{+16}_{-24}$ AU & $1.7^{+0.1}_{-0.3}$ & $0.19^{+0.03}_{-0.09}$ & $0.11^{+0.02}_{-0.03}$ AU & $1.2^{+0.1}_{-0.1}$\\
\enddata
\end{deluxetable*}

\section{DISCUSSION} \label{sec:disc}
\subsection{Possible Origin of the Surface Density Jump} \label{sec:jump}

A jump of the surface density between the envelope and the Keplerian disk around L1527 IRS has been found by our model fitting to the continuum visibility. In fact, such a density jump can be qualitatively confirmed in numerical simulations of disk evolution \citep{ma2010} and a similar density jump by a factor of $\sim 8$ is suggested to reproduce continuum emission arising from the disk around HH 212 \citep{le2014}. Furthermore it is important to note that the disk radius of L1527 IRS geometrically estimated with the density jump is fairly consistent with the radius kinematically estimated, suggesting that disks and envelopes around protostars may be not only kinematically but also geometrically distinguishable from the viewpoint of density contrast. The results also suggest that the density jump may be physically related to kinematical transition from infalling motions to Keplerian rotation. In this section, the possible origin of the surface density jump is quantitatively discussed. 

Surface density of the disk at the boundary $\sim 84$ AU can be calculated from our best-fit model to be $\Sigma _{\rm disk}= 0.42\ {\rm g}\, {\rm cm}^{-2}$ with Equation \ref{eq:sd} and the best-fit parameters shown in Table \ref{tab:mod}. This surface density is within the typical range derived for Class $\II$ YSO disks in Taurus \citep{an.wi2007} and Ophiuchus \citep{an2009,an2010}. On the other hand, volume density of the envelope at the boundary can be calculated from the best-fit model to be $\rho_{\rm env}= 1.0\times 10^{-16}\ {\rm g}\, {\rm cm}^{-3}$, which corresponds to the number density of $n_{\rm env}=2.5\times 10^{7}\ {\rm cm}^{-3}$. For embedded young stars in the Taurus-Auriga molecular cloud, a typical density distribution of protostellar envelopes can be described as $\sim (0.3-10)\times 10^{-13}\ {\rm g}\,{\rm cm}^{-3}\ (R/1\ {\rm AU})^{-3/2}$ \citep{ke1993}, which provides $(0.4-13)\times 10^{-16}\ {\rm g}\, {\rm cm}^{-3}$, corresponding to $(0.1-3.5)\times 10^{7}\ {\rm cm}^{-3}$, at the boundary, $R_{\rm out}$. This indicates that the envelope density in the best-fit model is also reasonable when compared with other observations. The envelope density is also consistent with that in \citet{to2013} at their boundary radius 125 AU.

Because the disk around L1527 IRS is considered to still grow with mass accretion from the envelope, as discussed in \citet{oh2014}, a possible origin of the density jump may be mass accretion from the envelope to the disk. 
First, density in the Keplerian disk is higher than that in the infalling envelope when radial motion in the disk is not as fast as that in the envelope, which is reasonable because gravity and centrifugal force are balanced in the disk while they are not balanced in the envelope. This difference of mass infall rates makes mass build up in the disk, resulting in disk growth. Secondly, to explain the factor $S_{\rm damp}=0.19$ quantitatively, we consider isothermal shock due to the mass accretion between the infalling envelope and the Keplerian disk. The gravity of a central protostar causes material in the envelope to infall dynamically and thus the material has radial infall velocity $u_{r} (R)$ as a function of radius. When $\rho _{\rm env}$, $\rho _{\rm disk}$, and $c_{s}$ indicate envelope density, disk density, and sound speed, respectively, the isothermal shock condition is $S_{\rm damp}=\rho _{\rm env}/\rho _{\rm disk}=c_{s}^{2}/u_{r}^{2}$. The sound speed at the boundary can be calculated from our best-fit model to be $c_{s}=0.39\ \kms$. Regarding the infall velocity, we assume $u_{r}$ to be the product of a constant coefficient $\alpha$ and free fall velocity, i.e., $u_{r}=\alpha \sqrt{2GM_{*}/R_{\rm out}}$, as was discussed by \citet{oh2014}. By using $M_{*}=0.45\ \Ms$ and $R_{\rm out}=84$ AU, this infall velocity can be calculated to be $3.1\alpha\ \kms$. In order to explain $S_{\rm damp}=0.19$, $\alpha$ should be $\sim 0.3$, which is consistent with the range of $\alpha$ (0.25-0.5) \citet{oh2014} found. This quantitative discussion suggests that mass accretion from the envelope to the disk is a possible origin of the density jump we found.

\subsection{Structures of the C$^{18}$O gas disk} 
\label{sec:gas}

The previous sections have discussed the structures of the disk and the envelope around L1527 IRS based on the continuum observations tracing the disk and the envelope. C$^{18}$O emission, on the other hand, also traces them, as was discussed in \citet{oh2014}. Because it can be reasonably assumed that gas and dust are well coupled and mixed in the protostellar phase in contrast to the T Tauri phase where gas and dust can be decoupled because of grain growth, it is important to examine whether the structures of the disk and the envelope revealed in the dust observations can also be valid for those traced in C$^{18}$O emission.

In order to answer the question mentioned above, in this section, models for C$^{18}$O are constructed based on the best-fit dust model derived from the fitting to the continuum visibility shown in Section \ref{sec:cv}, and models are compared with the C$^{18}$O observations. The models of C$^{18}$O have the same profiles of the surface density, temperature, and scale height as those of the best-fit dust model derived from the fitting to the continuum visibility shown in Section \ref{sec:cv}. The surface density profile has a jump at a radius of 84 AU, which is the boundary between the disk and the envelope, as was suggested by the best-fit dust model. In addition to these constraints on structures, C$^{18}$O models also require velocity fields, which cannot be constrained by the continuum observations. For velocity fields, we assume that C$^{18}$O models follow the radial profile of the rotation derived from the C$^{18}$O observations in Section \ref{sec:rp}, and that of infall introduced in Section \ref{sec:jump}. Note that we adopt 74 AU as the radius where the rotation profile has a break even though the geometrical boundary between the disk and the envelope is set at 84 AU in radius, as mentioned above. These geometrical and kinematical structures of the C$^{18}$O models are fixed in the following discussion. Importantly the C$^{18}$O models still depend on the fractional abundance of C$^{18}$O relative to H$_{2}$, $X({\rm C^{18}O})$, as discussed later.

In order to compare models and observations, C$^{18}$O data cubes are calculated from the models described above. It should be noted that the C$^{18}$O emission obviously traces more extended structures arising from outer parts of the envelope, as compared with the dust emission. Because structures of the disk and the envelope adopted for C$^{18}$O model are based on the continuum emission detected within a radius of $\sim 1\arcsec$ (see Figure \ref{fig:ci}), comparisons between C$^{18}$O models and observations should be made only within a radius of $1\arcsec$. According to the C$^{18}$O velocity channel maps shown in Figure \ref{fig:18c}, the C$^{18}$O emission arises within a radius of $\sim 1\arcsec$ when $|V_{\rm LSR}|>2.0\ \kms$, and only C$^{18}$O emission having these LSR velocities is discussed in comparisons between the models and the observations. When C$^{18}$O data cubes are calculated from the models, radiative transfer, including both dust and C$^{18}$O opacities, are also solved in 3D and velocity space, and then dust continuum emission is subtracted to derive the final model cubes. The model data cubes calculated from the radiative transfer are convolved with a Gaussian beam having the same major and minor axes, and orientation as the synthesized beam of our observations. A moment 0 map is made from this convolved data cube to compare with the observations. Even though visibilities are not compared between models and observations here, we can still judge how good each model is based on this comparison using moment 0 maps, as demonstrated in the model fitting for the continuum data in Section \ref{sec:fit}.

\begin{figure}[ht!]
\figurenum{9}
\epsscale{0.5}
\gridline{
\fig{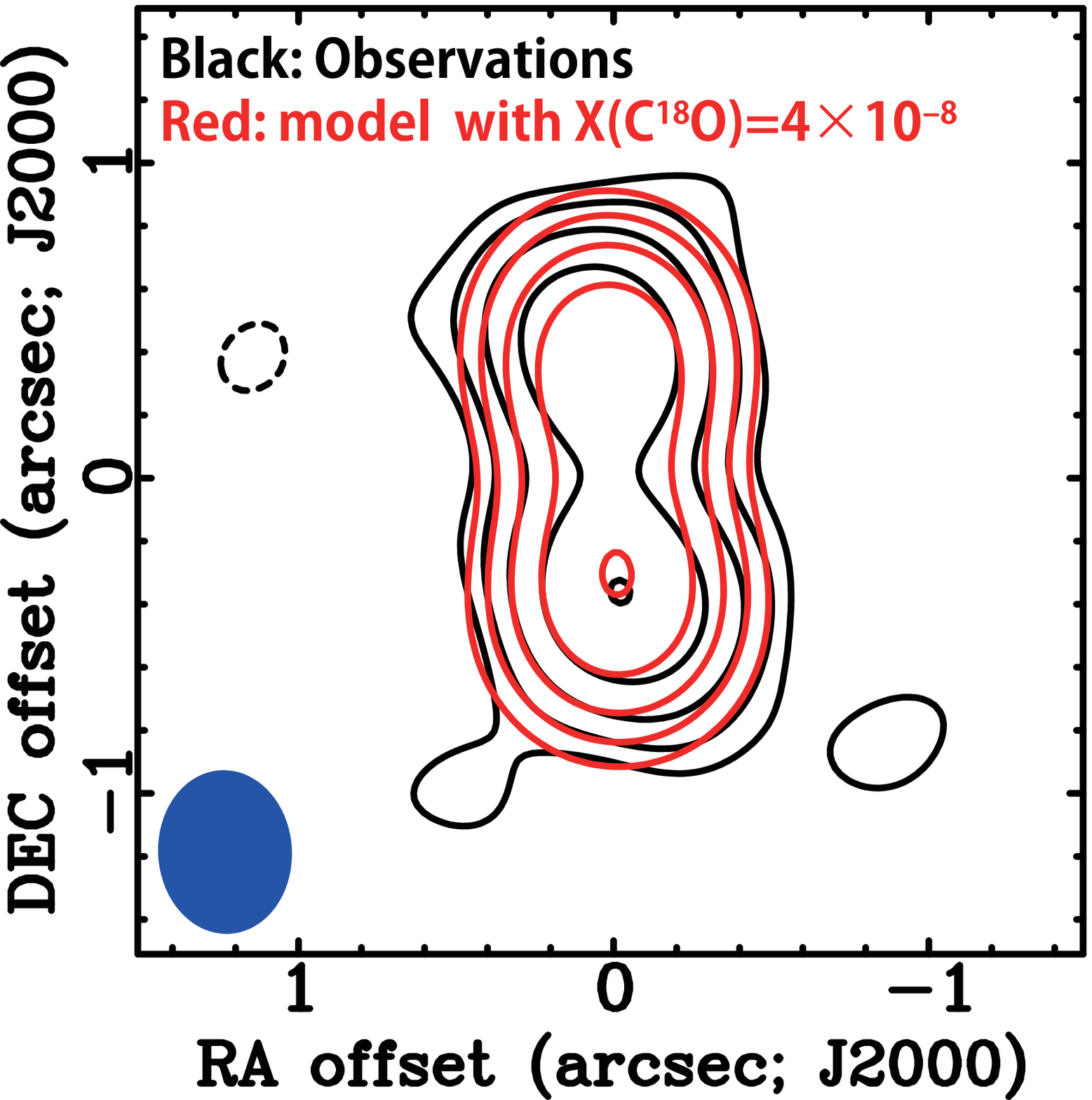}{0.25\textwidth}{(a)}
\fig{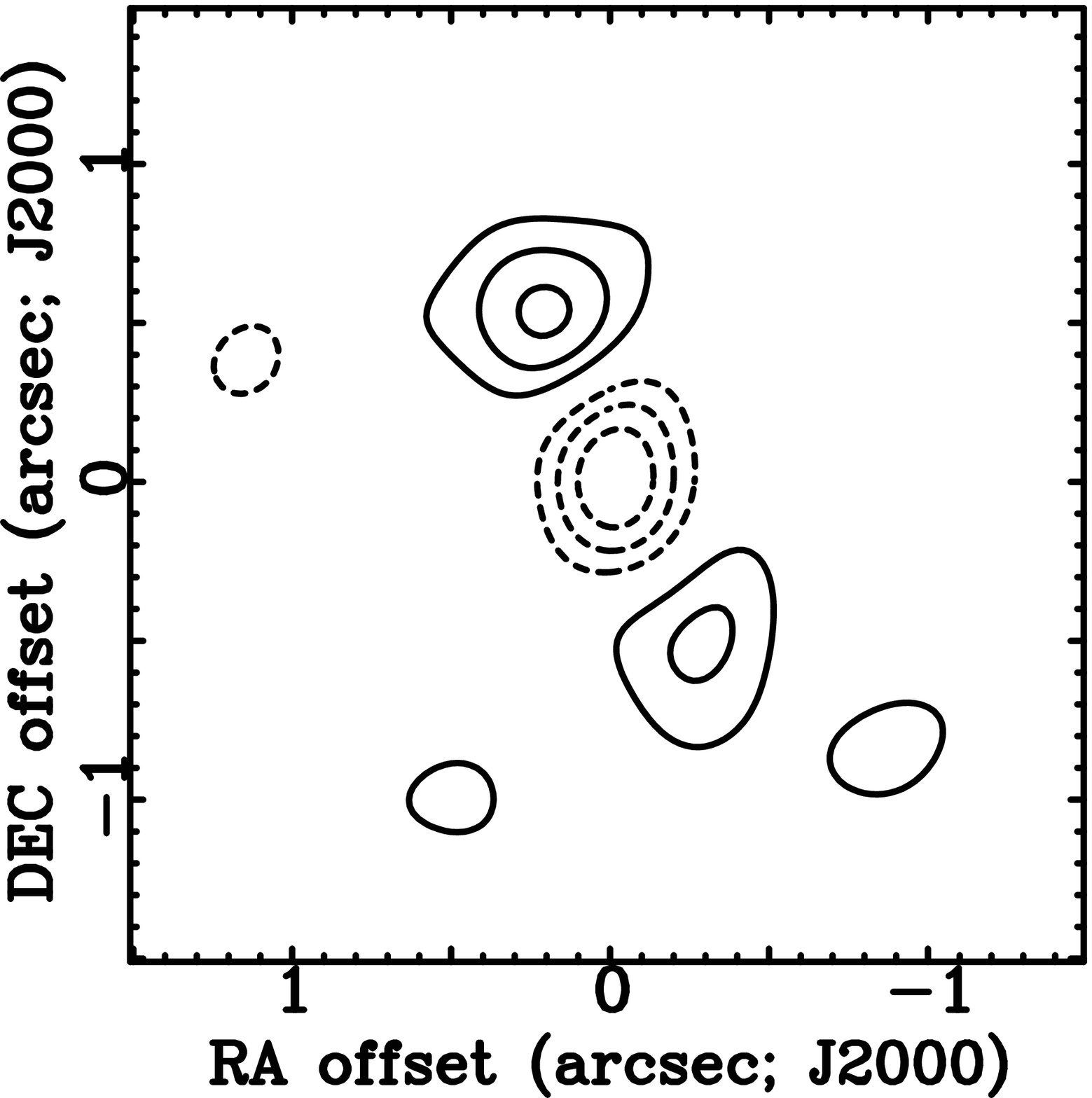}{0.25\textwidth}{(b)}
}
\caption{Comparison of C$^{18}$O moment 0 maps integrated over $|V|>2.0\ \kms$. (a) Observations in black contours and a model with $X({\rm C}^{18}{\rm O})=4\times 10^{-8}$ in red contours. (b) The residual obtained by subtracting the model from the observations. Contour levels are $-3,3,6,12,24,\dots \times \sigma$ in panel (a) and $-3,3,6,9,12,\dots \times \sigma$ in panel (b), where $1\sigma$ corresponds to $1.8\ \mJB \, \kms$. A blue filled ellipse at the bottom right corner denotes the ALMA synthesized beam; $0\farcs 50\times 0\farcs 40,\ {\rm P.A.}=3.1^{\circ}$.
\label{fig:cde}}
\end{figure}

Figure \ref{fig:cde}a shows a comparison of moment 0 maps between a model and the observations. In this model, a constant C$^{18}$O abundance of $4\times 10^{-8}$ is adopted as a nominal value. In this case, significant residuals at 9$\sigma$ level are left, as shown in Figure \ref{fig:cde}b. Inner regions show negative residuals while outer regions show positive residual, demonstrating that the model C$^{18}$O is too strong in inner regions while it is too weak in outer regions, as compared with the observations. Note that neither a higher nor a lower constant value than $4.0\times 10^{-8}$ of the C$^{18}$O abundance improves the model. For instance, the value in interstellar medium (ISM), $5.0\times 10^{-7}$ \citep{la1994,jo2005,wi.ro1994}, provides more negative residuals than Figure \ref{fig:cde}b. Although there are a couple of factors to change the C$^{18}$O intensity in the model, the abundance of C$^{18}$O is the only one that changes the intensity of C$^{18}$O if we still remain the same physical structures of the disk in the model, i.e., in order to make the C$^{18}$O intensity weaker in inner regions and stronger in outer regions in the model, the C$^{18}$O abundance might be lower in inner regions and higher in outer regions.
For instance, if the C$^{18}$O abundance in outer regions ($R>84$ AU) is the same as the one in ISM, $5.0\times 10^{-7}$ \citep{la1994,jo2005,wi.ro1994}, and the abundance in the inner regions ($R<84$ AU) decreases by a factor of $\sim 20$ due to the freeze-out of C$^{18}$O molecules, the observations can be explained by the model. T Tauri disks are, however, usually considered to have temperature profiles having higher temperature in inner regions \citep[e.g.,][]{an.wi2007}, and the disk around L1527 IRS is also considered to have such a temperature profile as \citet{to2013} suggested. With such a temperature profile, it would be difficult for the disk to have a lower C$^{18}$O abundance in the inner regions because of the molecular freeze-out.

\begin{figure}[ht!]
\figurenum{10}
\fig{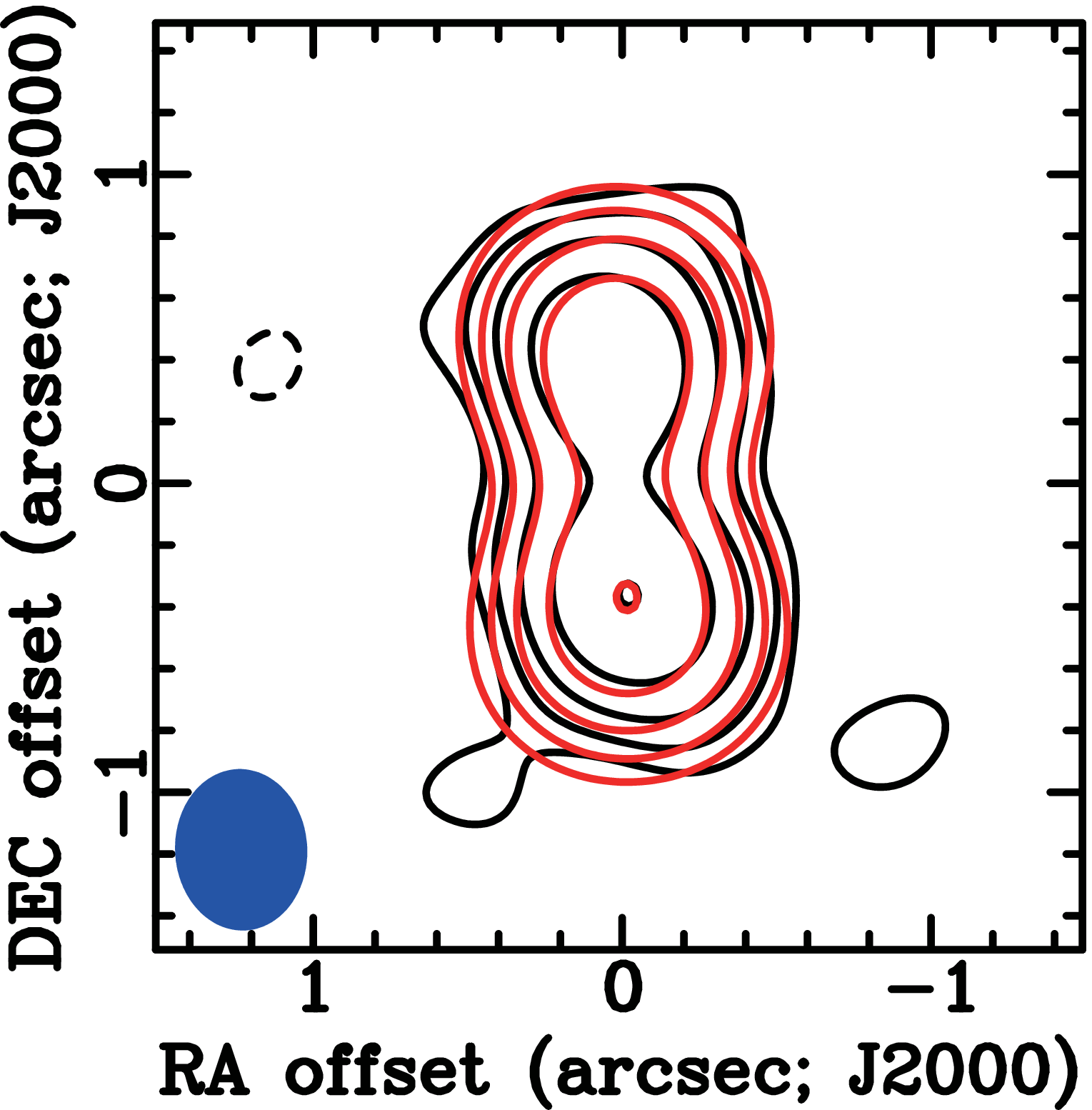}{0.23\textwidth}{(a)}
\fig{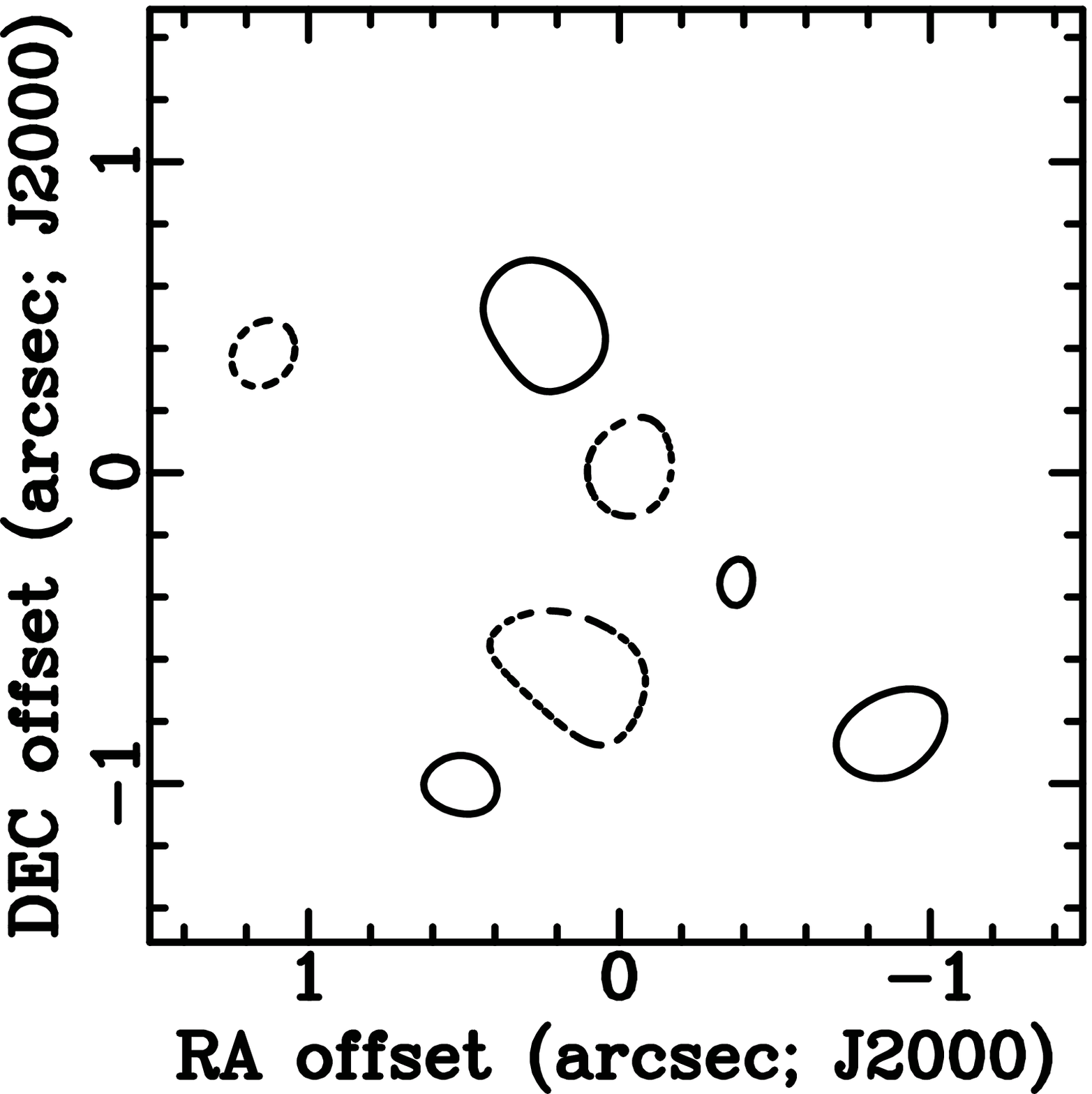}{0.23\textwidth}{(b)}
\caption{The same figures as Figure \ref{fig:cde} but the model has enhancement of C$^{18}$O abundance: $X({\rm C}^{18}{\rm O})=5.0\times 10^{-7}$ in 80-88 AU and $X({\rm C}^{18}{\rm O})=2.8\times 10^{-8}$ in the other regions.
\label{fig:cde2}}
\end{figure}

One possible C$^{18}$O abundance distribution that can explain the observations is the one with a local enhancement of the C$^{18}$O molecule, as has been suggested for the SO molecular abundance around L1527 IRS by \citet{oh2014}; they suggested that the SO abundance is locally enhanced around L1527 IRS because of accretion shocks making the dust temperature sufficiently high for SO molecules frozen out on dust grains to be desorbed.
An example of such a C$^{18}$O abundance distribution is the ISM abundance \citep[$5.0\times 10^{-7};$][]{la1994,jo2005,wi.ro1994} at 80$\leq R\leq $88 AU and a lower constant abundance of $2.8\times 10^{-8}$ elsewhere.
Figure \ref{fig:cde2} shows the comparison between the model with this C$^{18}$O abundance distribution and the observations, suggesting that the model with this C$^{18}$O abundance distribution can reproduce the observations with reasonably small residual.

The reason of the lower C$^{18}$O abundance in the inner disk region is not clear. According to the temperature distribution of L1527 IRS by \citet{to2013}, the midplane temperature becomes lower than 30 K at $r\gtrsim 100$ AU, sublimation temperature of CO on dense conditions \citep[10$^{8-12}$ cm$^{-3}$;][]{fu.ai2014}. Thus CO freeze-out could be present only at the outer region.
In general, CO molecules indeed could not be frozen-out so easily in protostellar disks as in T Tauri disks because surrounding envelopes heat up such embedded disks \citep{ha2015}, although the degree of this heating effect depends on a couple of factors, such as density distribution. 
There are two other possibilities to explain the observed C$^{18}$O abundance decrease in the inner disk region. One is that dust can be optically thick in inner regions and hides part of the C$^{18}$O emission as indicated by our best-fit model shown in Section \ref{sec:fit}. The other possibility is that chemistry on the warm and dense conditions. On such a condition CO can be converted into more complex molecules such as CO$_{2}$ and organic molecules. Indeed, recent CO observations of protoplanetary disks with ALMA have been reporting similar decrease of CO abundance, which is attributed to such a chemical effect \citep[e.g.,][]{sc2016}. The CO conversion is also reported in protostellar phases as well based on single-dish and interferometric observations \citep{an2016,fu2012,yi2012,al2010}.

Although it is difficult to give a strong constraint on the width of the local enhancement and the lower C$^{18}$O abundance in the frozen-out or converted region in models with the current observations, the density and temperature structures derived from the continuum observations can also reproduce the C$^{18}$O observations with a radial abundance profile of C$^{18}$O with a local enhancement like the one discussed above. Future observations at a higher angular resolution can give a better constraint on the C$^{18}$O radial abundance profile.

\section{CONCLUSIONS} \label{sec:conc}

We have observed the Class 0/I protostar L1527 IRS in the Taurus star-forming region with ALMA during its Cycle 1 in 220 GHz continuum and C$^{18}$O $J=2-1$ line emissions to probe the detailed structures of the disk and the envelope around L1527 IRS. 
The 220 GHz continuum emission spatially resolved with an angular resolution of $\sim 0\farcs5 \times 0\farcs 4$ shows a similar elongated structure in the north-south direction. Its deconvolved size is estimated from a 2D Gaussian fitting to be $\sim 0\farcs53 \times 0\farcs 15$, showing a significantly thinner structure than those previously reported on the same target. 
The C$^{18}$O $J=2-1$ emission overall shows an elongated structure in the north-south direction with its velocity gradient mainly along the same direction. The integrated intensity map shows a double peak with the central star located between the peaks, due to a continuum subtraction artifact. The elongation of the continuum as well as C$^{18}$O clearly indicates that these emissions trace the disk/envelope system around L1527 IRS and the velocity gradient along the elongation is naturally considered to be due to rotation of the system, as was previously suggested.

The radial profile of rotational velocity of the disk/envelope system obtained from the position-velocity diagram of the C$^{18}$O emission cutting along the major axis of the continuum emission was fitted with a double power-law, providing the best-fit result with a power-law index for the inner/higher-velocity ($p_{\rm in}$) of 0.50 and that for the outer/lower-velocity component ($p_{\rm out}$) of 1.22. This analysis clearly suggests the existence of a Keplerian disk around L1527 IRS, with a radius kinematically estimated to be $\sim 74 $ AU. The dynamical mass of the central protostar is estimated to be $\sim 0.45\ \Ms$.

In order to investigate structures of the disk/envelope system, $\chi ^{2}$ model fitting to the continuum visibility without any annulus averaging have been performed, revealing a density jump between the disk and the envelope, with a factor of $\sim 5$ higher density on the disk side. The disk radius geometrically identified as the density jump is consistent with the Keplerian disk radius kinematically estimated, suggesting that the density jump may be related to the kinematical transformation from infalling motions to Keplerian motions. One possible case to form such a density jump is isothermal shock due to mass accretion at the boundary between the envelope and the disk. If this is the case, to form the density jump with a factor of $\sim 5$ requires the infall velocity in the envelope to be $\sim 0.3$ times slower than the free fall velocity yielded by the central stellar mass. In addition to the density jump, it was found that the disk is roughly in hydrostatic equilibrium. The geometrical structures of the disk found from the $\chi^{2}$ model fitting to the continuum visibility can also reproduce the C$^{18}$O observations as well, if C$^{18}$O freeze-out, conversion, and localized desorption possibly occurring within $\sim 1\arcsec$ from the central star are taken into account.

\acknowledgments

This paper makes use of the following ALMA data: ADS/JAO.ALMA2012.1.00647.S (P.I. N. Ohashi). ALMA is a partnership of ESO (representing its member states), NSF (USA) and NINS (Japan), together with NRC (Canada), NSC and ASIAA (Taiwan), and KASI (Republic of Korea), in cooperation with the Republic of Chile. The Joint ALMA Observatory is operated by ESO, AUI/NRAO and NAOJ.
We thank all the ALMA staff making our observations successful. We also thank the anonymous referee, who gave us invaluable comments to improve the paper.
Data analysis were in part carried out on common use data analysis computer system at the Astronomy Data Center, ADC, of the National Astronomical Observatory of Japan.
S.T. acknowledges a grant from the Ministry of Science and Technology (MOST) of Taiwan
(MOST 102-2119-M-001-012-MY3), and JSPS KAKENHI Grant Number JP16H07086,
in support of this work.
Y.A. is supported by the Subaru Telescope Internship Program.


\vspace{5mm}
\facilities{ALMA}

\software{CASA, MIRIAD, IDL}

\bibliographystyle{aasjournal}
\bibliography{reference_aso}

\clearpage



\listofchanges

\end{document}